\documentclass[12pt,a4paper]{article}
\usepackage{amsmath}
\usepackage{amsfonts}
\usepackage{graphicx}
\usepackage{cite}
\usepackage{bbm}

\def\({\left(}
\def\){\right)}

\newcommand{\ket}[1]{|#1\rangle}
\newcommand{\bra}[1]{\langle #1 |}

\newcommand{\nn}{\nonumber}
\newcommand{\Eqn}[1]{&\hspace{-0.5em}#1\hspace{-0.5em}&}
\renewcommand{\[}{\begin{equation}}
\renewcommand{\]}{\end{equation}}
\newcommand{\eqb}{\begin{eqnarray}}
\newcommand{\eqe}{\end{eqnarray}}

\newcommand{\bbR}{{\mathbb R}}
\newcommand{\bbZ}{{\mathbb Z}}

\newcommand{\alg}[1]{\mathfrak{#1}}
\newcommand{\grp}[1]{\mathrm{#1}}

\newcommand{\bz}{{\bar{z}}}

\newcommand{\hlambda}{{\hat{\lambda}}}

\newcommand{\tn}{{\tilde{n}}}

\def\tr{{\rm tr}}

\def\bra{\langle}
\def\ket{\rangle}

\def\cdots {\cdot\cdot\cdot}
\def\pint#1 {- \!\!\!\!\!\!\!\! \,\int_{#1}}

\def\lb       {\left( }
\def\rb       {\right) }

\def\lbb     {\left[ }
\def\rbb      {\right] }
\def\comma      { \, , }
\def\period     { \, . }
\def\semiket#1  { \, #1 \, \rangle \, }
\def\del        {  \partial  }

\def\half       {\frac{1}{2}}
\def\abs#1      {  \, \vert #1 \vert \,   }
\def\Im#1    { \, {\rm Im } \, #1  }
\def\Re#1    { \, {\rm Re}  \, #1  }
\def\binom#1#2 { \vecii{ {}_{#1} }{\raisebox{.5ex}{$ {}^{#2} $}} }

\def\zbar   {\bar{z}}
\def\wbar   {\bar{w}}
\def\bbar   {\bar{b}}
\def\Zbar   {\bar{Z}}
\def\calG  {{\cal G}}
\def\calO   {{\cal O}}
\def\calR {{\cal R}}
\def\calM {{\cal M}}
\def\Chi {{\cal X}}

\def\tilmu   {\tilde{\mu}}
\def\tilM   {\tilde{M}}
\def\tilY   {\tilde{Y}}
\def\chat   { {\hat{c}} }
\def\bfm#1{\boldsymbol{#1}}
\newcommand{\One}{\mathbbm{1}}
\newcommand\bzero{\boldsymbol{0}}

\newcommand\brho{\boldsymbol{\rho}}

\def\vecii#1#2     {{ #1 \choose #2 }  }
\def\veciii#1#2#3  {\left(\begin{array}{c}#1\\#2\\#3\end{array}\right)}
\def\matrixii#1#2#3#4            {\Bigl( \begin{array}{cc}#1&#2\\#3&#4
                                   \end{array} \Bigr) }
\def\matrixiii#1#2#3#4#5#6#7#8#9 {\left(\begin{array}{ccc}#1&#2&#3\\
                                #4&#5&#6\\#7&#8&#9\end{array}\right)}

\renewcommand{\thesection}
{\arabic{section}.\hspace{-.5em}}
\renewcommand{\thesubsection}
{\arabic{section}.\arabic{subsection}.\hspace{-.5em}}
\renewcommand{\thesubsubsection}
{\arabic{section}.\arabic{subsection}.\arabic{subsubsection}.\hspace
                                                             {-.5em}}

\makeatletter
\renewcommand\section{
\@startsection{section}{3}{\z@}%
{-3.25ex\@plus -1ex \@minus -.2ex}%
{1.5ex \@plus .2ex}%
{\normalfont\normalsize\bfseries\mathversion{bold}}}
\renewcommand\subsection{
\@startsection{subsection}{3}{\z@}%
{-3.25ex\@plus -1ex \@minus -.2ex}%
{1.5ex \@plus .2ex}%
{\normalfont\normalsize\bfseries\mathversion{bold}}}
\renewcommand\subsubsection{
\@startsection{subsubsection}{3}{\z@}%
{-3.25ex\@plus -1ex \@minus -.2ex}%
{1.5ex \@plus .2ex}%
{\normalfont\normalsize\itshape}}
\makeatother

\makeatletter \@addtoreset{equation}{section} \makeatother
\renewcommand{\theequation}{\arabic{section}.\arabic{equation}}

\makeatletter
\renewcommand{\appendix}{
\renewcommand{\thesection}{\Alph{section}.\hspace{-.5em}}
\renewcommand{\thesubsection}
{\Alph{section}.\arabic{subsection}.\hspace{-.5em}}
\renewcommand{\thesubsubsection}
{\Alph{section}.\arabic{subsection}.\arabic{subsubsection}.\hspace
                                                           {-.5em}}
\@addtoreset{equation}{subsection}
\renewcommand{\theequation}{\Alph{section}.\arabic{equation}}
\setcounter{section}{0}}
\makeatother

\begin{document}
%
\def\papertitlepage{\baselineskip 3.5ex \thispagestyle{empty}}
\def\preprinumber#1#2#3#4{\hfill
\begin{minipage}{1.2in}
#1 \par\noindent #2 \par\noindent #3 \par\noindent #4
\end{minipage}}
\renewcommand{\thefootnote}{\fnsymbol{footnote}}
\newcounter{aff}
\renewcommand{\theaff}{\fnsymbol{aff}}
\newcommand{\affiliation}[1]{
\setcounter{aff}{#1} $\rule{0em}{1.2ex}^\theaff\hspace{-.4em}$}
%
%
\papertitlepage
\setcounter{page}{0}
\preprinumber{YITP-11-14}{TIT/HEP-610}{UTHEP-623}{} 
\vskip 1ex 
~
\vfill
\begin{center}
{\large\bf\mathversion{bold}
$g$-functions and gluon scattering amplitudes\\ at strong coupling}
\end{center}
\vfill
\baselineskip=3.5ex
\begin{center}
Yasuyuki Hatsuda\footnote[1]{\tt hatsuda@yukawa.kyoto-u.ac.jp}, 
Katsushi Ito\footnote[2]{\tt ito@th.phys.titech.ac.jp},
Kazuhiro Sakai\footnote[3]{\tt sakai@phys-h.keio.ac.jp} and
Yuji Satoh\footnote[4]{\tt ysatoh@het.ph.tsukuba.ac.jp}\\

{\small
\vskip 2ex
\affiliation{1}
{\it Yukawa Institute for Theoretical Physics, Kyoto University}\\
{\it Kyoto 606-8502, Japan}\\

\vskip 1ex
\affiliation{2}
{\it Department of Physics, Tokyo Institute of Technology}\\
{\it Tokyo 152-8551, Japan}\\

\vskip 1ex
\affiliation{3}
{\it Research and Education Center for Natural Sciences}\\
{\it and Hiyoshi Department of Physics, Keio University}\\
{\it Yokohama 223-8521, Japan}\\

\vskip 1ex
\affiliation{4}
{\it Institute of Physics, University of Tsukuba}\\
{\it Ibaraki 305-8571, Japan}

}
\end{center}
\vfill
%
\baselineskip=3.5ex
\begin{center} {\bf Abstract} \end{center}

We study gluon scattering amplitudes/Wilson loops in $\mathcal{N}=4$
super Yang--Mills theory at strong coupling by calculating the area of 
the minimal surfaces in $AdS_3$ based on the associated thermodynamic 
Bethe ansatz system. The remainder function of the amplitudes 
is computed by evaluating the free energy, the T- and Y-functions
of the homogeneous sine-Gordon model.
Using conformal field theory (CFT) perturbation, 
we examine the mass corrections to the free energy around 
the CFT point corresponding  to the regular polygonal Wilson loop.
Based on the relation between the T-functions 
and the $g$-functions, which measure the boundary entropy,  
we calculate corrections to the T- and Y-functions
as well as express them at the CFT point by the modular S-matrix.
We evaluate the remainder function around the CFT point
for 8 and 10-point amplitudes explicitly and compare
these analytic expressions with the 2-loop formulas.
The two rescaled remainder functions  show very similar
power series structures.

\vskip 2ex
\vspace*{\fill}
\noindent
February 2011
\setcounter{page}{0}
\newpage
\renewcommand{\thefootnote}{\arabic{footnote}}
\setcounter{footnote}{0}
\setcounter{section}{0}
\baselineskip = 3.5ex
\pagestyle{plain}
%

\section{Introduction}

It has been recognized 
that there exists an integrability structure in gluon scattering
amplitudes in ${\cal N}=4$
super Yang--Mills theory at strong coupling.
Gluon scattering amplitudes are dual to the Wilson loops made of
light-like segments\cite{Dr,Alday:2007hr}.
By using the AdS-CFT correspondence, the amplitudes at strong coupling 
are shown to be equal to the area of the minimal surface in AdS with the
same null polygonal boundary\cite{Alday:2007hr}.
For $n(\geq\!6)$-point amplitudes\cite{Dr2,Alday:2007he} 
they are shown to differ from
the Bern-Dixon-Smirnov (BDS) conjecture\cite{Bern:2005iz}.
The deviation from the BDS conjecture is called the remainder function,
which is a function of the cross-ratios of the external gluon momenta.

It has been found that the equations for
the minimal surface reduce to 
the Hitchin system and the area of the surface is determined by 
the Stokes data of its
solutions\cite{Alday:2009yn,Alday:2010vh,Alday:2009dv,Hatsuda:2010cc}.
The Stokes data and their cross-ratios 
obey the functional equations 
called the T-system \cite{Kuniba:2010ir}
and the Y-system\cite{Zamolodchikov:1991et}.
The area of the minimal surface is determined by 
the thermodynamic Bethe ansatz (TBA) 
equations \cite{Zamolodchikov:1989cf}
associated with the Y-system\cite{Alday:2010vh}.

Y-systems are closely related to two-dimensional integrable models.
For the 6-point gluon scattering amplitude, which corresponds to the 
minimal surface with  a hexagonal null boundary in $AdS_{5}$,
the related Y-system and the TBA equations are those of 
the $\bbZ_4$-symmetric integrable model\cite{Alday:2009dv}.
In our previous paper\cite{Hatsuda:2010vr}, we solved the TBA equations 
with chemical potential by the integrable perturbation of conformal
field theory (CFT) and evaluated the free energy.
In order to obtain the analytic form of the remainder function near
the CFT point corresponding  to the regular polygonal Wilson loop, 
it is moreover necessary to calculate small mass expansion of
the Y-functions,
which was determined numerically in \cite{Hatsuda:2010vr}.

A key observation for computing the Y-functions
analytically is as follows:
The TBA free energy is obtained from
the partition function on a cylinder
with the periodic boundary condition.
We can also consider the free energy on a cylinder with different
boundary conditions. 
Affleck and Ludwig \cite{Affleck:1991tk} introduced 
the $g$-functions from this free energy, which count
the ground state degeneracy of the system with boundaries
(the boundary entropy).
In \cite{Dorey:2004xk,Dorey:2005ak}, Dorey
et al.~studied the exact off-critical $g$-functions
for the purely elastic scattering theory and derived the integral
equation for them.
They evaluated the $g$-functions
using the integrable perturbation of the boundary CFT.
Remarkably, ratios of the $g$-functions obey
the same integral equations for the T-functions.
This enables us to obtain the analytic expression
for the Y-functions from the $g$-functions and determine
the analytic form of the remainder function near the CFT point.

In this paper, we study the remainder function for $2\tilde{n}$-point
gluon scattering amplitudes at strong coupling, which correspond to
minimal surfaces in $AdS_3$ 
with a $2\tilde{n}$-gonal light-like boundary. 
This corresponds to the case where the
gluon momenta are in $\mathbb{R}^{1,1}$.
In \cite{Hatsuda:2010cc} the related integrable system
was shown to be the homogeneous sine-Gordon model
(HSG) \cite{FernandezPousa:1996hi} 
with purely imaginary resonance parameters.
The relevant CFT is the generalized parafermions
${\rm SU}(\tilde{n}-2)_2/{\rm U}(1)^{\tilde{n}-2}$
\cite{Gepner:1987sm}.
In this paper we will study the boundary and bulk perturbation of 
the generalized parafermions and calculate
the ratios of the $g$-functions.
For the octagon ($\tilde{n}=4$) and the decagon ($\tilde{n}=5$), 
we will calculate the perturbative corrections
to the T-/Y-functions, the free energy and the remainder
function explicitly.
We compare these analytic expressions of the remainder function
with the 2-loop formulas proposed in 
\cite{DelDuca:2010zp,Heslop:2010kq,Gaiotto:2010fk} 
around the CFT limit. 

The above analytic results are important to understand
the structure and the momentum dependence of the amplitudes
at strong coupling exactly.
The purpose of this paper is to take a step toward this direction by
analyzing
the TBA system near the CFT point from the conformal perturbation 
theory (CPT). A point in our discussion is that not only the free energy
but also
the Y-functions can be discussed in this framework owing to the relation
between the T-/Y-functions and the $g$-functions.

This paper is organized as follows.
In section~2, we review the TBA system for
the minimal surfaces in $AdS_3$.
In section~3, we discuss the HSG model 
and the free energy for the integrable bulk perturbations.
In section~4, we study the HSG model from the CPT of
the generalized parafermions
and the relation between the T-functions and the $g$-functions.
In section~5, we investigate the small mass expansions of the
$g$-function and  the remainder 
function for the octagon. 
In section~6, we argue the corrections to the free energy,
the T-/Y-functions and the remainder function for the decagon.
In section~7, we compare the analytic expressions for the remainder
functions for the octagon and the decagon with  the 2-loop results.
In section~8, we give a summary and a discussion.
In Appendix~A, we present the high-temperature expansion of the
$g$-function for the Ising model in detail.
In Appendix~B, we study the CPT of
an ${\rm SU}(2)$ coset model with
fractional level.
In Appendix~C, we discuss the expansion of the Y-functions in the case
of complex masses.
In Appendix~D, we examine the structure of
the expansion of the remainder
functions at higher orders.

\section{Review of TBA system for minimal surfaces in $AdS_3$}

\subsection{Problem}

Gluon scattering amplitudes at strong coupling
are evaluated by the area of minimal surfaces
in $AdS_5$\cite{Alday:2007hr}.
The minimal surfaces have a polygonal boundary
located on the AdS boundary.
Each edge of the polygonal boundary
represents a gluon momentum.
Throughout this paper we consider the case where the
minimal surfaces stretch inside the maximal $AdS_3$ subspace.
In this case, the polygonal boundary is contained inside
$\bbR^{1,1}$ at the AdS boundary.
Namely, we consider amplitudes with 
all the gluon momenta being restricted in $\bbR^{1,1}$.
For general $2\tn$-point amplitudes,
we label the vertices of the polygon in the light-cone coordinates as
$x_{2k-1}=(x^+_k,x^-_k),\ x_{2k}=(x^+_{k+1},x^-_k),\ k=1,\ldots,\tn$
(with identification $x_{\tn+1}^\pm=x^\pm_1$).
The gluon momenta are then given by
$x_j-x_{j+1},\ j=1,\ldots 2\tn$ ($x_{2\tn+1}=x_1$).
These data completely fix the shape and
the area of the minimal surfaces.

As the gluon scattering amplitudes in ${\cal N}=4$ SYM are
infrared divergent, the area of the minimal surfaces are also divergent.
Nevertheless, the structure of the divergence has been well studied
and one can read off meaningful results
after specifying an appropriate regularization scheme.
The finite results are commonly analyzed
in the form of the remainder function,
namely the deviation of the area from
the conjecture of Bern, Dixon and Smirnov\cite{Bern:2005iz}.
It is a dual-conformally invariant quantity
and hence a function of cross-ratios of $x_j$.
The form of the function is of our central interest.

\subsection{T-/Y-system and TBA equations}

Despite the difficulty in analytically constructing
the minimal surfaces for general polygonal boundaries,
it turned out that the area can be computed
by solving TBA type integral equations.
This subsection is devoted to a brief summary
of the TBA equations and the associated T-/Y-system
for the minimal surfaces in $AdS_3$.
For details,
see \cite{Alday:2009yn,Alday:2010vh,Hatsuda:2010cc,Maldacena:2010kp}.

The standard procedure to analyze the minimal surfaces in $AdS_3$
is to consider the auxiliary linear problem
associated with the original nonlinear equations of motion.
For any minimal surface in $AdS_3$
one can consider 
$\grp{SU}(2)$ Hitchin equations\cite{Alday:2009yn},
which are a set of first order
linear differential equations defined
on the world-sheet with coordinates $z,\bz$.
To be precise, there appear two sets of Hitchin equations
corresponding to $\grp{SL}(2)_{\rm L}$, $\grp{SL}(2)_{\rm R}$
evolutions of $AdS_3$.
These are promoted to a one-parameter family of differential equations
with a spectral parameter $\zeta$,
where $\zeta=1$ and $\zeta=i$ correspond to the original
$\grp{SL}(2)_{\rm L}$ and $\grp{SL}(2)_{\rm R}$ cases, respectively.
The minimal surface is constructed as the product of the
solutions of these Hitchin equations.

When the minimal surface possesses a $2\tn$-gonal boundary,
solutions of the Hitchin equations exhibit the Stokes phenomena.
The world-sheet is divided,
with respect to the asymptotic behavior of solutions,
into $\tn$ angular regions called the Stokes sectors.
When moving from one sector to another at large $|z|$,
a solution shows drastic change in its asymptotic behavior.
This corresponds to moving from one cusp to
another cusp at the boundary of the minimal surface.
In each sector one can uniquely (up to normalization) choose
the ``small'' solution,
which shows the fastest decay for $|z|\to\infty$
among the solutions.
We let $s_j(z,\bz;\zeta)$ denote
such small solution in the $j$-th Stokes sector $(j=1,\ldots,\tn)$.
We are considering ($\zeta$-dependent)
$\grp{SU}(2)$ Hitchin equations,
and thus $s_j$'s are 2-component column vectors.
We fix the normalization of $s_j$ so that
\[\label{s_normalization}
\langle s_j,s_{j+1}\rangle := \det(s_j\, s_{j+1}) = 1.
\]

Since there are only two linearly independent solutions,
all $s_j$'s are expressed as linear combinations
of any two of them.
The coefficients of these linear combinations
are called the Stokes data.
The Stokes data are redundant
and one finds relations among them.
It turned out that such relations are concisely expressed
in the form of T-system, where Stokes data are identified
with T-functions\cite{Alday:2010vh}.
In the present case, the T-system reads
\eqb\label{Tsystem}
T^+_j T^-_j = 1+ T_{j-1}T_{j+1},
\eqe
where $j=1,\ldots,\tn-3$.
The T-functions are given by
\[\label{defT}
T_{2k+1}(\theta) = \bra s_{-k-1}, s_{k+1}\ket,\qquad 
T_{2k}(\theta) = \bra s_{-k-1} , s_{k}\ket^{+}
\]
for $T_j$ with $j=0,\ldots,\tn-2$
and the rest are set to be zero.
Here we introduced the new variable $\theta$ by
\[
\zeta=e^\theta
\]
and the convention $f^\pm := f(\theta \pm \pi i /2)$.
Note that Stokes data are by definition independent
of $z,\bz$ and depend only on $\zeta$.
In (\ref{defT}) there appear
$s_j$ with $j\le 0$,
which are defined by successively applying
the normalization condition (\ref{s_normalization}).
As $s_j$ and $s_{j+\tn}$ correspond to the same Stokes sector,
they are, as functions of $z,\bz$,
proportional to each other
\[
s_j \propto s_{j+\tn}.
\]
We see that $T_0=1$ and $T_{\tn-2}$ is equal to one of
the proportionality coefficients (monodromies).
The present T-system is
of the standard $A_{\tn-3}$ type.

While the T-functions completely characterize the shape of the
minimal surface, they contain unphysical gauge degrees of freedom.
Instead, one can consider gauge invariant quantities
called Y-functions. In this case, they are given by
\eqb\label{YinT}
 Y_j = T_{j-1} T_{j+1}.
\eqe
Note that $Y_j$ with $j=1,\ldots,\tn-3$ are in general
nontrivial functions while the rest are zero.
Y-functions correspond directly to the physical variables.
They are essentially the cross-ratios:
\eqb\label{YinChi}
 Y_{2k} = - \Chi_{-k,k,-k-1,k+1} \comma \quad
 Y_{2k+1} = - \Chi^{+}_{-k-1,k,-k-2,k+1} \comma
\eqe
where
\eqb
 \Chi_{ijkl} := \frac{\bra s_{i},s_{j}\ket \bra s_{k},s_{l}\ket }{
   \bra s_{i},s_{k}\ket \bra s_{j},s_{l}\ket } \comma
\eqe
and
\eqb\label{Chipm}
\Chi_{ijkl}(\zeta=1) = \frac{x^{+}_{ij}x^{+}_{kl}}{x^{+}_{ik}x^{+}_{jl}}
 =: \chi^{+}_{ijkl}
\comma \quad 
\Chi_{ijkl}(\zeta=i) = \frac{x^{-}_{ij}x^{-}_{kl}}{x^{-}_{ik}x^{-}_{jl}}
=: \chi^{-}_{ijkl},
\eqe
with $x_{ij}:=x_i-x_j$.
The indices in the cross-ratios $\chi^{\pm}_{ijkl}$
are labeled mod $\tn$, and the subscripts $\pm$ in $x_{ij}^{\pm}$
are space-time indices, not to be confused with the shift of $\theta$.

It follows from (\ref{Tsystem}) and (\ref{YinT}) that
the Y-functions satisfy the following Y-system
\eqb\label{Ysystem}
 Y^{+}_{j} Y^{-}_{j} = (1+Y_{j-1}) (1+Y_{j+1}) \comma
\eqe
where $j=1,\ldots,\tn-3$.
These are the main equations that
characterize the present Y-functions.
In addition,
the Y-functions for minimal surfaces in $AdS_3$
obey additional conditions.
One is 
\eqb\label{reality}
\overline{Y_j(\theta)} = Y_j(-\bar{\theta}),
\eqe
which follows from the reality condition of the minimal surfaces.
Another important condition is
the asymptotic behavior for small 
spectral parameter $\zeta = e^{\theta}\to 0$,
\eqb\label{Yasymp}
 \log Y_{2k} \sim \frac{Z_{2k}}{\zeta} \comma 
 \quad \log Y_{2k+1} \sim \frac{Z_{2k+1}}{i\zeta} \comma
\eqe
with moduli parameters $Z_{j}$.
This behavior is determined as follows: 
For large $|z|$, the small solutions behave
as $s_j\sim \exp(w/\zeta+\bar{w}\zeta)$,
where $w=\int^z\sqrt{p(z)}dz$
with $p(z)$ being the $(\tn-2)$-th degree polynomial
associated with the minimal surface\cite{Alday:2009yn}.
By the WKB analysis
one finds that the Y-functions behave as in (\ref{Yasymp})
with $Z_j= - \oint_{\gamma_j}\sqrt{p(z)}dz$.
Namely, $Z_j$ are period integrals
over cycles $\gamma_j$
of the Riemann surface $y^2=p(z)$.
$\gamma_j$ are taken in such a way that
each of them has nonzero intersection with the adjacent ones
$\gamma_{2k}\wedge\gamma_{2k-1}=\gamma_{2k}\wedge\gamma_{2k+1}=1$
\cite{Alday:2010vh}.
For later convenience, we rewrite $Z_j$
as ``mass'' parameters
\eqb
  m_{2k} = -2 Z_{2k} \comma \quad m_{2k+1} = -2 Z_{2k+1}/i \period
\eqe
The mass parameters are in general complex:
\eqb
 m_j = |m_j| e^{i \varphi_j} \period
\eqe
The other condition is the analyticity that
the shifted Y-functions
\eqb\label{Ytilde}
\tilY_j(\theta):= Y_j(\theta+i\varphi_j)
\eqe
are regular inside the strip
\eqb\label{Strip}
-\frac{\pi}{2}<{\rm Im}\,\theta<\frac{\pi}{2}.
\eqe

Incorporating these additional conditions,
one can transform the Y-system relations (\ref{Ysystem})
into the following TBA type integral equations
\eqb\label{TBAeqs}
 \log \tilY_{j}(\theta) = -|m_{j}| \cosh \theta
  + K_{j,j-1} \ast \log (1+\tilY_{j-1})
  +  K_{j,j+1} \ast \log (1+\tilY_{j+1})
\eqe
for $|\varphi_{j}- \varphi_{j\pm 1}| < \pi/2$, 
where
\eqb\label{Kjj}
 K_{jj'}(\theta) :=
\frac{1}{2\pi} \frac{1}{\cosh(\theta + i\varphi_{j} - i \varphi_{j'})}
\eqe
and $f\ast g := \int f(\theta -\theta') g(\theta') d\theta'$.
The free energy associated with the TBA system is 
expressed by the solutions of the above TBA equations
as
\eqb\label{Afree}
 -F = A_{ \rm free}
 = \sum_{j=1}^{\tn-3} \int_{-\infty}^{\infty} \frac{d\theta}{2\pi}
   |m_{j}| \cosh \theta \log\bigl(1+\tilY_{j}(\theta) \bigr).
\eqe
This gives the main contribution
to the area of the minimal surfaces.

\subsection{Remainder function}

The TBA equations (\ref{TBAeqs}) completely determine the Y-functions.
The area of the minimal surface and the reminder function
are then computed by using these Y-functions.
The remainder function is defined as
\eqb
  R = A -A_{ \rm div} -A_{ \rm BDS},
\eqe
where $A$ is the area of the minimal surface,
$A_{\rm div}$ and $A_{\rm BDS}$ are
respectively the divergent term and the finite term
read from the BDS conjecture\cite{Alday:2009yn}.
The formulas for the $2\tn$-point amplitudes are different
for odd $\tn$ and even $\tn$. We describe them separately below.

\subsubsection{Case with odd $\tn$}

In this case, the remainder function reads
\eqb
R =  A_{ \rm sinh} + A_{ \rm periods} + \Delta A_{ \rm BDS}  \period
\eqe
The first term is 
\eqb\label{Asinh}
 && A_{ \rm sinh} = A_{ \rm free} + (\tn-2) A_{ \rm sinh}^{\tn=3} \comma
\eqe 
where $ A_{ \rm sinh}^{\tn=3} = 7\pi/12$
is the value for the hexagon solution,
which necessarily becomes equivalent to the regular hexagon solution. 

The second term is given by \cite{Alday:2009yn} 
\eqb\label{Apodd2}
A_{ \rm periods} = i \sum_{r=1}^{(\tn-3)/2} 
  \bigl( \bar{w}_{r}^{e} w^{m,r} - w_{r}^{e} \bar{w}^{m,r}\bigr) \comma
\eqe
where $w^{e}_{r}=\oint_{\gamma_r^e}\sqrt{p(z)}dz$ and 
$w^{m,r}=\oint_{\gamma^{m,r}}\sqrt{p(z)}dz$ are
the periods for the cycles 
with the canonical intersection form
$\gamma_{r}^{e} \wedge \gamma^{m,s} = \delta_{r}^{s}$.
In terms of the periods $Z_j$,
which correspond to cycles
with nontrivial intersection form
$\theta^{jk}=\gamma_j\wedge\gamma_k$
(see \cite{Alday:2010vh}),
it is written as
\eqb
 A_{ \rm periods} = -i w_{jk} Z^{j}\Zbar^{k}.
\eqe
Here, $w_{jk}$ is the inverse of the intersection form $\theta^{jk}$.

The third term is given by \cite{Alday:2009yn}
\eqb\label{DeltaBDS1}
\Delta A_{ \rm BDS} = A_{\mbox{\scriptsize BDS-like}} - A_{ \rm BDS} 
= \frac{1}{4} \sum_{i,j = 1}^{\tn}
\log\frac{ c_{i,j}^{+} }{ c_{i,j+1}^{+} } 
\log \frac{ c_{i-1,j}^{-} }{c_{i,j}^{-} } \period
\eqe
$A_{\mbox{\scriptsize BDS-like}}$ is a finite term
left after subtracting $A_{\rm div}$ from the area
of a certain reference region.
$c^{\pm}_{i,j}$ are cross-ratios formed by nearest neighbor distances
only.
For example, 
when $(j-i)>0$ is odd,
\eqb\label{defcij}
 c^{\pm}_{i,j} =
 \frac{ x^{\pm}_{i+2,i+1} x^{\pm}_{i+4,i+3} \cdots x^{\pm}_{j,i} }{
 x^{\pm}_{i+1,i} x^{\pm}_{i+3,i+2} \cdots x^{\pm}_{j,j-1} },
\eqe
where $x_i,x_{i+1},\ldots,x_j$ successively appear in the expression.
When $(j-i)>0$ is even the path goes
along the opposite side:
$x_i\to x_{i-1}\to\cdots\to x_j$.
By definition, $c^\pm_{i,j} = c^\pm_{j,i}$.
The indices of $c_{i,j}^{\pm}$
are labeled mod $\tn$.
We also define $c^\pm_{i,j} = 1$ for $|i-j| \leq 1$.
For the $AdS_{3}$ kinematics, one can
set the coordinates of the cusps
so that $c^\pm_{i,j} > 0$. 
Note that
one can express $c^\pm_{i,j}$ in terms of Y-functions
at special values by using (\ref{YinChi}), (\ref{Chipm}).

\subsubsection{Case with even $\tn$}

In this case,
\eqb\label{eq:Reven}
R =  A_{ \rm sinh} + A_{ \rm periods} 
  + A_{ \rm extra} + \Delta A_{ \rm BDS}  \period
\eqe
The first term $A_{ \rm sinh}$ is the same as in (\ref{Asinh}).
The second term is 
\cite{Alday:2009yn}
\eqb
 A_{ \rm periods} = i \sum_{r=2}^{(\tn-2)/2} 
 \bigl( \bar{w}_{r}^{e} w^{m,r} - w_{r}^{e} \bar{w}^{m,r}\bigr) \period
\eqe
The third term is given by
\eqb
A_{ \rm extra}
 = -\frac{1}{2} (w_{\rm s} +\wbar_{\rm s}) \log \gamma_{1}^{R}
  + \frac{1}{2i} (w_{\rm s} -\wbar_{\rm s}) \log \gamma_{1}^{L} \comma
\eqe
where
\eqb
 e^{w_{\rm s} +\wbar_{\rm s}}
 = - \frac{x^{+}_{23}x^{+}_{45} \cdots x^{+}_{\tn,1}}{
  x^{+}_{12}x^{+}_{34} \cdots x^{+}_{\tn-1,\tn}} \comma
  \quad
  e^{(w_{\rm s} -\wbar_{\rm s})/i}
 = - \frac{x^{-}_{23}x^{-}_{45} \cdots x^{-}_{\tn,1}}{
  x^{-}_{12}x^{-}_{34} \cdots x^{-}_{\tn-1,\tn}} \comma
\eqe
and 
\eqb\label{gammaLR}
 \gamma_{1}^{L} = \gamma_{1} (\zeta=1) \comma \quad 
 \gamma_{1}^{R} = \gamma_{1} (\zeta=i) \comma
\eqe
with 
\eqb\label{eq:gamma1T1}
\gamma_{1} (\zeta) = T_{1} (\theta+\pi i) \period
\eqe
Here, $e^{w_{\rm s} +\wbar_{\rm s}}$,
$e^{(w_{\rm s} -\wbar_{\rm s})/i}$ are given by
$(T_{\tn-2})^{(-1)^{\tn/2+1}}$
at $\theta =-\pi i/2$ and $0$, respectively.
The relation  between $\gamma_{1}(\zeta)$ and $T_{1}$
follows from the definition of the Stokes data.
We see that $A_{\rm extra}$ is expressed in terms of T-functions.
The T-functions are expressed in terms of Y-functions,
as we will see in section 4.

The last term is similar to (\ref{DeltaBDS1}),
\eqb
\Delta A_{ \rm BDS} = A_{\mbox{\scriptsize BDS-like}} - A_{ \rm BDS} 
= \frac{1}{4} \sum_{i,j = 1}^{\tn'}
\log\frac{ \chat_{i,j}^{+} }{ \chat_{i,j+1}^{+} } 
\log \frac{ \chat_{i-1,j}^{-} }{\chat_{i,j}^{+} } \comma
\eqe
where $\tn'=\tn+1$ and $\chat^{\pm}_{ij}$ are defined by 
$c_{ij}$ for the $2(\tn+1)$-point case by the double soft limit
$x_{\tn+1}^{\pm} \to x_{1}^{\pm}$. The explicit form is given in 
\cite{Maldacena:2010kp}.

\section{Integrable models and CFT perturbation}

As we saw in the last section, the problem of computing
gluon scattering amplitudes at strong coupling
is governed by TBA type integral equations.
It was found \cite{Hatsuda:2010cc} that the above TBA equations
for null-polygonal minimal surfaces in $AdS_3$ are
identified with those of the homogeneous sine-Gordon (HSG) models
\cite{FernandezPousa:1996hi}.
In particular, when the resonance parameters are trivial,
the HSG models admit the description of
the bulk and boundary conformal perturbation theory (CPT).
This allows us to analytically solve the TBA equations
near the CFT point in the form of
high-temperature (or small mass) expansion.
Moreover, in some special cases
the TBA equations are also identified with those of
other integrable theories.
In these cases, the CFT analysis becomes simpler.
Furthermore, 
from the results for the trivial resonance parameters
one is able to analyze the case of nontrivial resonance parameters,
as is discussed later on.
In this section we review
those integrable models associated with the TBA equations
and the high-temperature expansion of the free energy.

\subsection{Homogeneous sine-Gordon model}

The HSG models are 
obtained as integrable perturbations \cite{IntPerturbation}
of the coset $ \grp{G}_k/\grp{U}(1)^{ r_{\grp{G}} }$ CFTs
(generalized parafermion CFTs)\cite{Gepner:1987sm,parafermions}.
Here, $k$ is the level and $r_{\grp{G}}$
is the rank of  a compact Lie group $\grp{G}$.
For the present purpose
we focus on the case with $\grp{G}=\grp{SU}(n)$.
The classical action is 
\eqb
 S= k \Bigl( S_{\rm gWZW} -  \int d^2x \, V(g) \Bigr) \comma
\eqe
where $S_{\rm gWZW} $ is the corresponding gauged WZW action,
$g $ is an element of $\grp{G}$, and
\eqb
V(g) = \frac{\mu^2}{4\pi} \tr(\Lambda_+ g^\dagger \Lambda_- g ) \period
\eqe
$\Lambda_\pm = i \bfm{\lambda}_\pm\cdot \bfm{h}$ are elements 
of the Cartan subalgebra $\alg{h}$ of the Lie algebra
$\alg{g}=\alg{su}(n)$, which are parametrized
by two $(n-1)$-dimensional vectors 
\eqb\label{lambdapm}
\bfm{\lambda}_\pm 
= \sum_{i=1}^{n-1} \tilmu_i e^{\pm \sigma_i} \bfm{\lambda}_i
\period
\eqe 
Here $\bfm{\lambda}_i$ are the fundamental weights,
the parameters $\sigma_i$ describe the resonance
when they are real and
together with a bare overall mass scale $\mu$, 
dimensionless  parameters $\tilmu_i$ give the semi-classical masses of
the solitons
\eqb
 \mu_i^a = \mu \tilmu_i \rho_a \comma \qquad 
  \rho_a = \frac{\sin \frac{\pi a}{k} }{\sin \frac{\pi}{k}}
  \quad  (a = 1, ..., k-1) \period
\eqe
The potential  is identified with a linear combination of 
weight-zero adjoint fields. 
One can also rewrite the potential $V(g)$ as
\eqb
 V(g) = \frac{\mu^2}{4\pi}
\sum_{i,j=1}^{n-1} \mu_{ij}^2 \Gamma_{ij}(g) \comma
\eqe
with
\eqb
 \mu^2_{ij} = \tilmu_i \tilmu_j e^{\sigma_i -\sigma_j} \comma \quad
 \Gamma_{ij}(g) = - \tr\Bigl( (\bfm{\lambda}_i\cdot \bfm{h})  g^\dagger  
  (\bfm{\lambda}_j\cdot \bfm{h}) g \Bigr) \comma
\eqe
so that it becomes manifest that
each mass scale $\mu_{ij}$ is turned on by the field $\Gamma_{ij}$.

Let us now consider the quantum theory.
For the reason explained in the next subsection,
we set $\sigma_i=0$.
In the quantum theory, the adjoint fields have conformal dimensions
\eqb
\Delta = \bar{\Delta}
= \frac{n}{k+n}.
\eqe
Thus, as in the case of the sine-Gordon model\cite{Zamolodchikov:1995xk},
on dimensional grounds the potential is renormalized
as
\eqb
V \sim  \sum_{i,j=1}^{n-1}
 (M^2 \tilM_i \tilM_j)^{1-(\Delta+\bar{\Delta})/2} 
 \Bigl[ \Gamma_{ij}(g) \Bigr]_{\calR}.
\eqe
As in the bare case, $M$ is the physical mass scale and $\tilM_i$ are
dimensionless parameters, in terms of which
the physical masses are expressed as
\eqb
 M_i^a =  M_i \rho_a  \comma \quad M_i = M \tilM_i \period  
\eqe
Replacing $\Bigl[ \Gamma_{ij}(g) \Bigr]_{\calR}$ with
the weight-zero adjoint operators,
the action in the quantum theory may be given by
\eqb\label{CPTaction}
S = k S_{\rm gWZW} + \lambda \int d^2x \, 
 \Phi_{\bfm{\lambda}, \bar{\bfm{\lambda}}}
 \comma
\eqe
where
\eqb\label{perturbingop}
\Phi_{\bfm{\lambda}, \bar{\bfm{\lambda}}}
&=&  (\bfm{\lambda})^l (\bar{\bfm{\lambda}})^{\bar{l}} \phi_{l,\bar{l}}
\nn \\
& = & \sum_{i,j=1}^{n-1}
(\tilM_i \tilM_j)^{1-(\Delta+\bar{\Delta})/2}  
   (\bfm{\hlambda}_j)^l (\bfm{\hlambda}_i)^{\bar{l}} \phi_{l,\bar{l}}
   \period
\eqe
Here, 
$\phi_{l,\bar{l}}$ are the adjoint operators  which
transform as the $l$-th ($\bar{l}$-th) element of $\alg{h}$
under the left (right) transformation.
We have set $\sigma_i = 0$
and hence $\bfm{\lambda} = \bar{\bfm{\lambda}}$.
One can decompose them as
\eqb
\bfm{\lambda} =
\bar{\bfm{\lambda}} = 
\sum_{j=1}^{n-1} \tilM_j^{1-(\Delta+\bar{\Delta})/2} \bfm{\hlambda}_j
   \comma
\eqe
where each basis vector $\bfm{\hlambda}_j$ corresponds
to the deformation along which $\tilM_j$ varies.
Classically $\bfm{\hlambda}_j$
coincide with the fundamental weights $\bfm{\lambda}_j$,
but quantum mechanically
they are functions of the ratios of $\tilM_j$.
The coupling constant is related to the mass scale $M$
as\footnote{
Throughout this paper, we consider the perturbation with
a negative coupling constant $\lambda$.}
\eqb\label{lambdaM}
\lambda = - \kappa_n M^{2-(\Delta+\bar{\Delta})} \period
\eqe
The proportionality constant $\kappa_n$ is computed explicitly
for some simple cases, as we will see later.

We normalize the adjoint operators as
\eqb
\Big\bra \phi_{l,\bar{l}}(z) \phi_{l',\bar{l}'}(0) \Big\ket =
\delta_{ll'} \delta_{\bar{l}\bar{l}'} \frac{1}{|z|^{4\Delta} } \comma
\eqe
from which
\eqb\label{Phi2pt}
 \Bigl\langle \Phi_{ \bfm{\lambda}, \bar{ \bfm{\lambda} } }(z)
  \Phi_{ \bfm{\lambda},\bar{ \bfm{\lambda} }}(0) \Bigr\rangle
= \frac{G^{2}}{ |z|^{\frac{4n}{n+2} } } \comma \qquad
   G(\tilM_{j}) := \sum_{i,j=1}^{n-1}
   \tilM_{i}^{\frac{2}{n+2}}F_{ij}  \tilM_{j}^{\frac{2}{n+2}}
\comma
\eqe
where
\eqb
F_{ij} = \bfm{\hlambda}_i \cdot \bfm{\hlambda}_j.
\eqe
Note that in the classical limit $F_{ij}$ coincides with
the inverse of the Cartan matrix.

\subsection{Correspondence with minimal surfaces in $AdS_3$}

Let us consider the $\grp{SU}(n)$ HSG model at level $k=2$.
For the case with $k=2$, we can drop the index $a$
in $M_i^a$ and absorb $\rho_a$ into $M$.
In \cite{Miramontes:1999hx}, the exact S-matrix
of HSG models associated with simply laced $\alg{g}$
is proposed. For the $\grp{SU}(n)_2$ HSG model,
the S-matrix reads \cite{CastroAlvaredo:2000nr}
\eqb\label{HSGS}
 S_{jk}(\theta) &=&
  (-1)^{\delta_{jk}} 
  \Bigl[ c_j \tanh \half(\theta +\sigma_j-\sigma_k
                        -\frac{\pi}{2}i)\Bigr]^{I_{jk}}
\eqe
for $j,k = 1, ..., n-1$, 
where $\delta_{jk}$ is the Kronecker delta, $c_j= \pm 1$, 
and $I_{jk}$ is the 
incidence matrix for $\alg{su}(n)$.
Given the S-matrix, one can write down
the TBA equations for the HSG models
\cite{CastroAlvaredo:1999em,CastroAlvaredo:2000nr,Dorey:2004qc}
as
\eqb\label{HSGTBAeqs}
 \epsilon_j(\theta) = M_jL \cosh \theta
  - K_{jk} \ast \log (1+e^{-\epsilon_k}),
\eqe
where $L$ denotes the inverse temperature and the kernel is given by
\[
K_{jk}=i\frac{\partial}{\partial\theta}\ln S_{jk}.
\]
As usual, the pseudo energies $\epsilon_j(\theta)$
are related to the Y-functions as
\[
\tilY_j=e^{-\epsilon_j}.
\]
It was found in \cite{Hatsuda:2010cc}
that the above TBA equations coincide with
those for the minimal surfaces in $AdS_3$ (\ref{TBAeqs})
under the identification
\[
n = \tn-2,\qquad
M_jL = |m_j|,\qquad
\sigma_j = i\varphi_j.
\]

Note that the TBA equations for the minimal surfaces
with complex masses ($\varphi_j\ne 0$)
correspond to those for
the HSG model with purely imaginary resonance parameters.
Physical interpretation of such resonance parameters
in the HSG model is not quite clear at present.
In addition, when $\sigma_j-\sigma_k$ are nonzero,
the boundary Yang--Baxter equations \cite{Cherednik:1985vs}
are not satisfied 
and the boundary factorizable scattering
may not be well-defined.
These are the reasons why we have set $\sigma_j=0$
in the last subsection.
Also, with this restriction one can make
full use of the known results about
the high-temperature expansion, as we see in the following.
We will discuss how to incorporate $\sigma_{j}$ in later sections.

\subsection{Bulk perturbation of free energy}

The free energy of the model on a space
of length $R \gg 1$ with temperature $1/L$ 
(in the $L$-channel)
gives the ground state energy of the model on a space 
of length $L$ (in the $R$-channel)\cite{Zamolodchikov:1989cf}. 
The free energy around the CFT point is
then given by evaluating the ground state energy
of the perturbed CFT on a cylinder of circumference $L$
with small coupling constant $\lambda$.
From the action (\ref{CPTaction}),
the conformal perturbation theory (CPT)
gives an expansion of the free energy\footnote{
We have rescaled the free energy as $L^2F/R \to F$.}
\eqb\label{CPTF}
-F \, =  \, A_{\rm free} \Eqn{=} \frac{\pi}{6} c_n + f_n^{\rm bulk}
+(2\pi)^2 \sum_{k=1}^{\infty} \frac{(-\lambda)^{k}}{k!} 
   \Bigl( \frac{2\pi}{L}\Bigr)^{2(\Delta-1)k} \nn  \\
 && \ \, \times  
\int \Big\bra \Phi_{\bfm{\lambda},\bar{\bfm{\lambda}}}(z_{1},\zbar_{1}) 
\cdots \Phi_{\bfm{\lambda},\bar{\bfm{\lambda}}}(z_{k},\zbar_{k}) 
 \Big\ket_{\rm connected} 
\prod_{i=2}^{k} (z_{i}\zbar_{i})^{\Delta-1} dz^{2}_{2} \cdots dz^{2}_{k}
  \nn\\
&&\\
 \Eqn{:=} \frac{\pi}{6} c_n +
 f_n^{\rm bulk}+ \sum_{k=2}^\infty f_{n}^{(k)} l^{\frac{4k}{n+2}}. \nn
\eqe
Here, $l=ML$ denotes the scale parameter,
$c_n$ is the central charge and $f_n^{\rm bulk}$ is the bulk term.
We have set $z_{1}=1$ by using translational invariance.
We have used the fact that
the coupling constant $\lambda$ has mass dimension
$2-2\Delta=\frac{4}{n+2}$ and is proportional to
$l^{\frac{4}{n+2}}$.
We have also used the fact that the one-point function
vanishes.
The central charge
of the coset $\grp{SU}(n)_2/\grp{U}(1)^{n-1}$ is
\[
c_{n} = \frac{n(n-1)}{n+2}.
\]

The bulk term $f^{\rm bulk}_n$ may be obtained
as a generalization of the results in the literature
\cite{Zamolodchikov:1989cf, Klassen:1990dx, Zamolodchikov:1991vh, 
Zamolodchikov:1991vx}.
For odd $n$, following the procedure in
\cite{Zamolodchikov:1989cf, Klassen:1990dx} (see also \cite{Mussardo})
one arrives at the expression\footnote{In the course of
the derivation in the references,
$\Delta$ is assumed to be sufficiently small, 
so that the terms in the summation in (\ref{CPTF}) are smaller than
$l^2$.
As this condition is not valid in our case with $\Delta = n/(n+2)$,
one needs appropriate modification for a rigorous derivation.
We have checked for $n=3$ that the expression (\ref{eq:Bnodd})
is in good agreement with numerical results. }
\eqb\label{eq:Bnodd}
 f_n^{\rm bulk}
= \frac{1}{4} l^2\sum_{i,j=1}^{n-1}
\tilM_i (I^{-1})_{ij} \tilM_j \period
\eqe
Note that the incidence matrix $I_{ij}$ has the inverse for odd $n$.
For even $n$, following the analysis in
\cite{Zamolodchikov:1991vx} one obtains
\eqb
 f_n^{\rm bulk}= \frac{1}{(n+2)\pi} Q^2 \cdot  l^{2} \log l \comma
\eqe
where
\eqb\label{Qdef}
Q := \sum_{j=0}^{n/2-1} (-)^j \tilM_{2j+1} \period
\eqe
(\ref{eq:Bnodd})--(\ref{Qdef}) reproduce the known results
in the case of a single nonzero mass
\cite{Zamolodchikov:1991vh, Zamolodchikov:1991vx, Zamolodchikov:1991vg}.

The coefficients $f_{n}^{(k)}$ are obtained by
computing CFT correlation functions.
Since we already know the two-point function, we find the 
lowest correction from  CPT:
\eqb\label{fn2}
f_{n}^{(2)} = \frac{\pi}{6}  \kappa_{n}^{2}  G^{2}  C_{n}^{(2)}
\comma \quad C_{n}^{(2)} = 3 (2\pi)^{\frac{2(n-2)}{n+2}}
\gamma^{2}\Bigl( \frac{n}{n+2}\Bigr) \gamma\Bigl( \frac{2-n}{n+2}\Bigr)
\period
\eqe
Here, $G$ is given in (\ref{Phi2pt}) and we have used 
\eqb
 \int d^{2}z \, |z|^{2a} |1-z|^{2b}
= \pi \gamma(1+a) \gamma(1+b) \gamma(-1-a-b)
 \period
\eqe
In section~6, we present the explicit expression
in the case of $n=3$ as an illustration.

\subsection{Single mass cases}

The TBA equations (\ref{HSGTBAeqs}) contain $n-1$ mass parameters.
When some of them are set to be zero,
the corresponding massless pseudo energies $\epsilon_j(\theta)$
become finite constants for $L\to\infty$,
while massive ones become infinite\cite{Zamolodchikov:1991vh}.
This gives rise to the appearance of CFT with
a different central charge.
Nevertheless, one finds little difference
concerning the small mass expansion:
When one turns on/off some of the masses,
only the constant term changes discontinuously
and the other expansion coefficients changes continuously.
This can be checked numerically.

For HSG models with general masses,
the precise form of the perturbing
operator (\ref{perturbingop}), as a function of masses,
has not yet been well understood.
On the other hand, when only one mass parameter is nonzero,
the CFT in the small mass limit becomes simple
and the perturbing operator as well as
the exact coupling--mass ratio is known.
This allows us to explicitly compute
the expansion coefficients at low orders.
Below we present a list of such cases
and collect useful facts in performing
the small mass expansion.\footnote{
The reader is referred to \cite{Ravanini:1992fi}
for a more extensive classification of
conformal perturbations associated with TBA equations.}
Remarkably, combining the cases in
subsections 3.4.1 and 3.4.3,
we are able to analyze a case of two general masses,
which we will discuss in section~6.

\subsubsection{Integrable perturbation of unitary minimal model}

Let us first consider the case
\[
M_1 = M,\quad \mbox{others}=0.
\]
The TBA equations (\ref{HSGTBAeqs}) in this case
are identified with those of
the (RSOS)${}_{n}$ scattering theory\cite{Zamolodchikov:1991vh,Itoyama:1990pv}.
The (RSOS)${}_{n}$ scattering theory is regarded as 
the massive perturbation of unitary minimal model $\calM_{n+1,n+2}$ 
by the primary field $\Phi_{1,3}$
with dimension $\Delta = \bar{\Delta} = \frac{n}{n+2}$.
The action for the (RSOS)${}_{n}$ scattering theory is
\eqb
 S_{{\rm (RSOS)}_{n}} = S_{\calM_{n+1,n+2}}
+ \lambda_{\rm RSOS} \int d^2 x \, \Phi_{1,3}
 \comma
\eqe
where the relevant operator is normalized as
\eqb\label{Phi132pt}
 \Big\bra \Phi_{1,3}(z) \Phi_{1,3}(0)\Big\ket
= \frac{1}{|z|^{4\Delta}} \comma
\eqe
and the coupling $\lambda_{\rm RSOS}$ is related to the mass $M_{1}$
as \cite{Zamolodchikov:1995xk}
\eqb
\lambda_{\rm RSOS} = - \kappa_{n}^{\rm RSOS} M_{1}^{2(1-\Delta)} \comma
\eqe
with 
\eqb\label{eq:kappaRSOS}
 \kappa_{n}^{\rm RSOS} = \frac{1}{\pi} \frac{(n+2)^{2}}{n(2n+1)} 
 \lbb \gamma\Bigl( \frac{3(n+1)}{n+2}\Bigl)
      \gamma\Bigl( \frac{n+1}{n+2} \Bigr)\rbb^{\half}
 \lbb \frac{\sqrt{\pi}
 \Gamma(\frac{n+2}{2})}{2\Gamma(\frac{n+1}{2})} \rbb^{\frac{4}{n+2}}
 \comma
\eqe
and $\gamma(x) = \Gamma(x)/\Gamma(1-x)$.

Taking into account the normalizations (\ref{Phi2pt}),
(\ref{Phi132pt}), one finds that
\eqb
 \lambda G(\tilM_{i}) \Big\vert_{\tilM_{i}
=0 (i\neq 1)} = \lambda_{\rm RSOS} 
 \comma
\eqe
and hence
\eqb
 \lambda = - \kappa_n M^{\frac{4}{n+2}},\qquad
 \kappa_{n} = \frac{\kappa_{n}^{\rm RSOS} }{F_{11}}.
\eqe
%

\subsubsection{Integrable perturbation of
             unitary SU(2) diagonal coset}

The above case is generalized to the cases
\[
M_k = M,\quad \mbox{others}=0,\quad\mbox{for}\quad k=1,\ldots,n-1.
\]
It is known that the TBA equations (\ref{HSGTBAeqs}) in these cases
describe the system obtained as integrable perturbation of
$(\grp{SU(2)}_k\times\grp{SU(2)}_{n-k})/\grp{SU(2)}_n$
coset CFT \cite{Zamolodchikov:1991vg,Ravanini:1992fi}.
The perturbing operator is the primary field
$\phi_{1,1,3}$, which corresponds to
the branching of the product of two trivial
representations into the adjoint representation.
This operator has conformal dimension
$\Delta=\bar\Delta=\frac{n}{n+2}$.
The exact coupling--mass ratio in these cases
is found in \cite{Fateev:1993av}.

\subsubsection{Integrable perturbation of non-unitary
             SU(2) diagonal coset}

Let us next consider the cases where
\[
M_k=M_{n-k}=M,\quad \mbox{others} = 0,\quad (k=1,\ldots,n-1)
\]
with $n$ being odd.
This kind of configuration is invariant under the $\bbZ_2$
outer automorphism of the $A_{n-1}$ Dynkin diagram.
In this case, one can regard the TBA equations
as those corresponding to the tadpole diagram
$T_{(n-1)/2}\,(\,=A_{n-1}/\bbZ_2)$
with a single mass parameter being turned on.
It is known that
they describe the system obtained as integrable perturbation of
non-unitary
$(\grp{SU(2)}_k\times\grp{SU(2)}_{n/2-k-1})/\grp{SU(2)}_{n/2-1}$
coset CFT \cite{Ravanini:1992fi}.
The perturbing operator is $\phi_{1,1,3}$
with dimension $\Delta=\bar\Delta=\frac{n-2}{n+2}$.

In particular, the case with $k=1$, namely
\[
M_1=M_{n-1}=M,\quad \mbox{others} = 0
\]
corresponds to integrable perturbation of
the non-unitary minimal models ${\cal M}_{n,n+2}$.
The case with $n=3$ and $k=1$
will be used in the analysis for the decagon in section~6.

\section{Conformal perturbation of $g$- and T-functions}

As we saw in the previous section, the relation between the
free energy in the $L$-channel and the ground state energy 
in the $R$-channel allows one to derive an 
expansion of the free energy near the CFT limit.
Such an expansion is studied for the 6-point amplitudes
with the $AdS_{5}$
kinematics in \cite{Hatsuda:2010vr}. To obtain the full expression 
of the amplitudes, one further needs the expansion of the
cross-ratios or the Y-/T-functions. A key observation 
\cite{Bazhanov:1994ft,Dorey:1999cj}
for this purpose is the relationship between the T-functions
and the $g$-functions 
\cite{Affleck:1991tk}.
In this section, we discuss the conformal perturbation of the 
T-functions of the HSG model associated with the coset 
$\grp{SU}(n)_{2}/\grp{U}(1)^{n-1}$. 
We follow \cite{Dorey:1999cj,Dorey:2005ak}
where the perturbation of the $g$- and T-functions in 
the ADET purely elastic scattering theories is discussed.
The discussion below holds for general $n$
with trivial formal monodromy (i.e., $T_n=1$)
and will be applied to concrete examples in the following
sections.\footnote{For even $n$, one can consider the cases with
non-trivial $T_n$, where the discussion may need some modifications.
Such cases appear for $2n+4\ge 12$,
which we hope to discuss elsewhere.}

\subsection{$g$-functions in homogeneous sine-Gordon model}

We start  by considering the partition function
$Z_{\bra \alpha \vert \alpha \ket} [L,R]$ on a cylinder of circumference $L$,
length $R$, and boundary conditions of type $\alpha$ at both 
ends. It is expanded by  the eigenvalues of the circle Hamiltonian
$H^{\rm circ}(M,L)$ as
\eqb
Z_{\bra \alpha \vert \alpha \ket} [L,R] 
= \bra \alpha \vert \, e^{-R H^{\rm circ}(M,L)} \vert \alpha \ket
= \sum_{p=0}^{\infty} \Bigl( \calG_{\vert \alpha \ket}^{(p)}(l) \Bigr)^2 
  e^{-R E_{p}^{\rm circ}(M,L)} \comma
\eqe
where $M$ is the mass scale defined through $M_{k} = \tilM_{k}M$, 
\eqb\label{calGp}
\calG_{\vert \alpha \ket}^{(p)}(l) 
 = \frac{\bra \alpha \vert \psi_p \ket}{ \bra \psi_p \vert \psi_p \ket^{1/2} }
\eqe
with $ \vert \psi_p \ket$ being the eigenstates of the Hamiltonian, and $l=ML$.
The $g$-function  is then defined by 
subtracting  the linear term
 in $L$ from $\calG_{\vert \alpha \ket}^{(0)}(l) $:
\eqb
\log g_{\vert \alpha \ket}(l)  = \log \calG_{\vert \alpha \ket}^{(0)}(l) + f_{\vert \alpha \ket} L
\period
\eqe
$f_{\vert \alpha \ket}$ is the constant boundary contribution
to the ground state energy $E_0^{\rm strip}$ of the $L$-channel Hamiltonian
$H^{\rm strip}_{\bra \alpha \vert \alpha \ket}(R)$.
The $g$-function counts the ground state degeneracy and is known to
decrease along the renormalization flow \cite{Affleck:1991tk}. It is also a subleading
contribution to the partition function for large $R$.

To consider the $g$-functions in the HSG model, we assume that
the HSG model admits
an integrable  generalization with  boundaries, and the boundary scattering
amplitudes are also diagonal. 
In this section, we also set 
the parity breaking parameters $\sigma_{j}=0$
as discussed in section 3.
We will discuss how to incorporate $\sigma_{j}$ in later sections.

In the presence of boundaries, one has the boundary reflection
factors $R_j(\theta)$, which are constrained by unitarity and crossing-unitarity
\cite{Fring:1993mp,Ghoshal:1993tm},
\eqb\label{Runicross}
R_j(\theta) R_{j}(-\theta) = 1 \comma \quad
R_j(\theta) R_{\bar{\jmath}}(\theta- i \pi) = S_{jj}(2\theta) \comma
\eqe
where the anti-particles are the same as the particles in our case, $\bar{\jmath} = j$.
The boundary Yang-Baxter equations are
indeed satisfied for $\sigma_{jk}:=\sigma_{j} -\sigma_{k} =0$.
In general,  reflection factors also have to satisfy the boundary bootstrap equations.
They are, however,  trivial in our case, since the bulk S-matrix (\ref{HSGS})
with $\sigma_{jk} =0 $  
does not have poles in the physical strip  
$ 0 \leq {\rm Im}\,\theta \leq \pi$.
The form of the constraints (\ref{Runicross}) shows that a set of reflection
factors $R_j(\theta)$ generates another $R'_j =R_j/Z_j$, if $Z_j(\theta)$ are 
a solution to one-index versions of the bulk unitary and crossing-unitary 
equations \cite{Sasaki:1993xr},
\eqb\label{Zj}
Z_j(\theta) Z_j(-\theta) = 1 \comma \quad 
 Z_j(\theta) Z_{\bar{\jmath}}(\theta-i\pi) = 1 \period
\eqe
In particular, we use a solution,
\eqb
Z_j^{\vert k,C \ket}(\theta) 
:= \Bigl( (1+C)_{\theta}(1-C)_{\theta}  \Bigr)^{\delta_{jk}}
\comma
\eqe
in the following, where
\eqb
(x)_{\theta} := \frac{\sinh \half(\theta + i\frac{\pi }{2}x)}{
\sinh \half(\theta - i\frac{\pi }{2}x)} \period
\eqe

A boundary $ \alpha$ is then associated with a set of the reflection factors 
$R^{\vert \alpha \ket }_{j}(\theta)$. It turns out that the $g$-function corresponding to
$\vert \alpha \ket$  
satisfies the following integral equation 
\cite{Dorey:2004xk,Dorey:2005ak,Pozsgay:2010tv,Woynarovich:2010wt},
\eqb\label{geq}
\log g_{\vert \alpha \ket}(l) \Eqn{=}  \log C_{\vert\alpha \ket } + \Sigma(l)   \\
&& \quad + \ \frac{1}{4}
\sum_{j=1}^{n-1}
\int_{\mathbb{R}} d\theta
\left( \phi^{\vert \alpha \ket }_{j}(\theta){-}
\delta(\theta){-}2\phi_{jj}(2\theta)\right)
  \log \left(1+e^{-\epsilon_j(\theta)}\right)
\period \nn
\eqe
Here, $C_{\vert \alpha \ket} $ is a symmetry factor associated with 
the vacuum degeneracy at infinite $l $. $ \phi^{\vert \alpha \ket }_{j}$
and $\phi_{jk}$ are given by the boundary and bulk S-matrices as
\eqb
\phi^{\vert \alpha \ket }_{j} (\theta) 
= \frac{1}{\pi i} \del_\theta \log R^{\vert \alpha \ket }_{j}(\theta)
\comma \quad
\phi_{jk} (\theta) 
= \frac{1}{2\pi i} \del_\theta \log S_{jk}(\theta)  \period
\eqe
$\Sigma(l)$ is a certain boundary-condition  independent term,
precise form of which is irrelevant for our purpose.
For details, see  \cite{Dorey:2004xk,Dorey:2005ak}.

\subsection{Relation between $g$- and T-functions}
In the conformal limit,  boundary conditions are labeled by primary fields,
and hence a boundary $ \alpha  $ may be
specified by the corresponding primary field.
To describe the relation between the $g$- and the T-functions,
we then consider the reflection factors
corresponding to the boundary condition labeled by the identity 
operator, $ R_j^{\vert \One \ket}(\theta)$, together with  the deformed ones,
\eqb\label{RkC}
R_j^{\vert k,C\ket }(\theta) 
= \frac{R_j^{\vert \One \ket}(\theta)}{Z_j^{\vert k,C \ket}(\theta)}
\period
\eqe
$R_j^{\vert \One \ket}$  is expected to be minimal, namely, having
the smallest number of poles and zeros. We assume the existence
of $R_j^{\vert \One \ket}$. 
These reflection factors give the $g$-functions $g_{\vert \One \ket}$ and 
$g_{\vert k,C \ket}$ through (\ref{geq}), the ratio of which satisfies
an integral equation,
\eqb\label{geq2}
&&\log\lb \frac{g_{\vert k,C \ket}(l)}{g_{\vert \One \ket}(l)} 
 \frac{C_{\vert \One \ket}}{C_{\vert k,C \ket}} \rb
 \\
&& \qquad  \qquad 
= \int \frac{d\theta}{4\pi} \lb \frac{1}{\cosh(\theta+i\frac{\pi}{2}C)}
 + \frac{1}{\cosh(\theta-i\frac{\pi}{2}C)} \rb
 \log\Bigr(1+Y_k(\theta) \Bigr)
  \period  \nn 
\eqe
We notice that 
almost only the information of the deforming  factors $Z_j^{\vert k,C \ket}$
has remained. 

On the other hand, from the T-system  (\ref{Tsystem})
and the relation between the
T- and Y-functions  (\ref{YinT}),  the T-functions obey
the integral equation,
\footnote{
For even $n$, an appropriate gauge  has to be chosen.}
\eqb\label{Teq2}
\log T_k(\theta) = -\nu_k \cosh \theta + K\ast\log(1+Y_k) \period
\eqe
Here,  the kernel is given by (\ref{Kjj}) with
$\varphi_{j} = \varphi_{j'} =0$: $K=1/(2\pi \cosh \theta)$.
$\nu_{k}$ specify the asymptotic behavior of $T_{k}(\theta)$ and
are related to the mass term of the TBA equations as 
\eqb
m_{k} = \nu_{k-1} + \nu_{k+1} \period
\eqe
Comparing (\ref{geq2}) and  (\ref{Teq2}), and
using $T_k(\theta) = T_k(-\theta) $, 
one finds that
\eqb
 \log\lb \frac{g_{\vert k,C \ket}(l)}{g_{\vert \One \ket}(l)} 
\frac{C_{\vert \One \ket}}{C_{\vert k,C \ket}} \rb
= \nu_k \cos\Bigr( \frac{\pi}{2} C\Bigr) + \log T_k\Bigl(i\frac{\pi}{2}C\Bigr) \period
\eqe
Since $C_{\vert \alpha \ket}$ is associated with the vacuum degeneracy, which 
may be determined by the symmetry, we expect 
$ C_{\vert \One \ket} = C_{\vert k,C \ket}$. 
Assuming this and subtracting the linear term in $ l \propto \nu_k $, we 
arrive at an important formula,
\eqb\label{gT}
\frac{\calG_{\vert k,C \ket }^{(0)}}{\calG_{\vert \One \ket}^{(0)}}
 = T_k\Bigl(i\frac{\pi}{2}C\Bigr) \period
\eqe

\subsection{Expansion of T-functions}

Using (\ref{gT}) and the conformal perturbation, one can derive 
an expansion of $T_{k}(\theta)$ near the CFT limit. 
To this end, we first note that
the conformal perturbation gives the expansion of 
the $g$-functions \cite{Dorey:1999cj,Dorey:2005ak},
\eqb\label{Gexpansion}
\log \calG^{(0)}_{\vert \One \ket}(\lambda,L) &=&\sum_{q =0}^{\infty}g_q^{\vert \One \ket}
(\lambda L^{2- 2\Delta})^q 
\comma \nn \\
\log \calG^{(0)}_{\vert \alpha \ket} (\lambda, \mu ,L) 
&=& \sum_{p,q =0}^{\infty}g_{p,q}^{\vert \alpha \ket} 
(\mu L^{1-\Delta})^p (\lambda L^{2- 2\Delta})^q
\comma
\eqe
where $\mu$ is the coupling of the boundary perturbation, and 
we have assumed that the dimension of the boundary 
perturbing operator is  $\Delta=n/(n+2)$. 
Following the argument in \cite{Dorey:2005ak}, we also assume the
relation between the boundary coupling and the deformation parameter,
\eqb\label{muC}
\mu = \mu_{0} \cos\Bigl( \frac{\pi}{n+2} C\Bigr) \comma
\eqe
where $\mu_{0}$ is some constant.  
Together with (\ref{gT}), this relates the boundary coupling $\mu$
to the argument of  $T_{k}(\theta)$. 
(\ref{muC}) also means that the boundary coupling vanishes
at $C=(n+2)/2$. In this case, the boundary conditions become conformal
and are described by (linear combinations of) the Cardy boundary states.
The conformal perturbation with only the bulk coupling $\lambda$ turned on
then gives 
\cite{Dorey:1999cj,Dorey:2005ak} 
\eqb\label{Gexpansion2}
\log \calG_{\vert \alpha \ket} (\lambda, L) 
:= \log \bra \alpha \vert \Omega \ket
= \log g_{\vert \alpha \ket}
+ \lambda d_1^{\vert \alpha  \ket} L^{2(1-\Delta)} + \cdots \period
\eqe
Here,  $\vert \Omega \ket$ is the full ground state, 
\eqb
d_1^{\vert \alpha \ket}  = - \frac{1}{2(2\pi)^{1-2\Delta}} 
\frac{ g^\Phi_{\vert \alpha \ket}}{ g_{\vert \alpha \ket}}
B(1-2\Delta, \Delta) \comma
\eqe
and 
$B(a,b) = \Gamma(a)\Gamma(b)/\Gamma(a+b)$ is the Euler beta function. 
We have also introduced
\eqb
g_{\vert \alpha \ket} := \bra \alpha \vert 0 \ket \comma  \quad
g_{\vert \alpha \ket}^{\Phi}: = \bra \alpha \vert \Phi \ket \comma 
\eqe
with $\vert 0 \ket$ and $\vert \Phi \ket$  being
the CFT vacuum state and  the state corresponding to the bulk perturbing field 
$\Phi =\Phi_{\bfm{\lambda}, \bar{\bfm{\lambda}}}$,
respectively.
$\calG_{\vert \alpha \ket} $ is different from 
$\calG_{\vert \alpha \ket}^{(0)} $ in (\ref{calGp}) by a normalization factor
$\bra \Omega \vert \Omega \ket^{1/2}$,
but this factor  is cancelled in the formula (\ref{gT}).

On the T-function side,
we first note that the Y-functions  for the $2(n+2)$-point amplitudes
have  the periodicity,
\eqb\label{Yperiod}
Y_k\Bigl(\theta + i\pi\frac{n+2}{2}\Bigr) = Y_{n-k}(\theta)  
\period
\eqe
$Y_k$ are also analytic for finite $\theta$, and even functions of $\theta$,
$Y_{k}(\theta) = Y_{k}(-\theta)$. These properties are common to
the T-functions, which  leads to  the expansion \cite{Zamolodchikov:1991et},
\eqb\label{Texpansion}
T_k(\theta) &=& \sum_{p=0}^{\infty}  c_k^{(p)}(l) \cosh\Bigl(\frac{2p}{n+2} \theta \Bigr)
\comma
\eqe
with
\eqb
c_{k}^{(2q)}(l) =   c_{n-k}^{(2q)}(l)  \comma \quad 
c_{k}^{(2q+1)}(l) =  -  c_{n-k}^{(2q+1)}(l)  \quad (q \in \bbZ_{\geq 0}) \period
\eqe
For small $l$, the coefficients behave as $c_{k}^{(p)}(l) \sim l^{(1-\Delta)p}$, 
since the Y- and T-functions show plateaus for $ -\log (1/l) \ll \theta \ll \log(1/l)$. 
The conformal perturbation (\ref{Gexpansion}) further
suggests that $c_{k}^{(p)}(l)$ are   expanded as
\eqb\label{cexpansion}
c_{k}^{(p)}(l) = \sum_{q=0}^{\infty} c_{k}^{(p,2q)} l^{(1-\Delta)(p+2q)} \period
\eqe
(See Appendix C and D for details.)
The Y-functions have a similar expansion.
Similar double expansions have been discussed also for 
other TBA systems \cite{Dorey:1997rb,Hatsuda:2010vr}.
Substituting this expansion into the Y-system  (\ref{Ysystem}),
one then finds that some lower coefficients  vanish. 
In term of the T-functions, the result reads 
\eqb\label{Tccoeff}
& c_k^{(0)}(l) = c_k^{(0,0)} + o\bigl(l^{3(1-\Delta)}\bigr) \comma
\quad  c_k^{(1)}(l) =  0+ o\bigl(l^{3(1-\Delta)}\bigr) \comma  & \nn\\
& c_k^{(2)}(l) =  c_k^{(2,0)} l^{2(1-\Delta)}  + o\bigl(l^{3(1-\Delta)}\bigr) \period  &
\eqe

Now, from  (\ref{gT}),   (\ref{Gexpansion2}), 
(\ref{Texpansion}) and (\ref{Tccoeff}) with $C = (n+2)/2$,  
the expansion of $T_{k}(\theta)$ is determined.
Comparing  both sides of (\ref{gT}), 
one first notices that 
the two expansions (\ref{Gexpansion}) and (\ref{Tccoeff})
are consistent with each other, once $g_{1,0}^{\vert \alpha \ket} =0$ is
taken into account, meaning that
the one-point functions of the boundary perturbing operator 
vanish in unitary theories.
Furthermore,
\eqb\label{Tcexpansion}
T_k(\theta) =  c_k^{(0,0)} + c_k^{(2,0)} l^{2(1-\Delta)} 
\cosh\Bigl( \frac{4\theta}{n+2} \Bigr)
+ \calO(l^{3(1-\Delta)}) \comma
\eqe
where
\eqb
c_k^{(0,0)} =  \frac{ g_{\vert k \ket}}{ 
g_{\vert \One \ket}} \comma
\quad
c_k^{(2,0)} =   \frac{ g_{\vert k \ket}}{ 
g_{\vert \One \ket}}
\bigl(d_1^{ \vert k \ket }
- d_1^{ \vert \One \ket } \bigr) \kappa_{n} \comma
\eqe
and  we have used  (\ref{lambdaM}).
We have also set $g_{\vert k,C \ket} =: g_{\vert k\ket}$,
$g_{\vert k,C \ket}^{\Phi} =: g_{\vert k\ket}^{\Phi}$ and
$d_{1}^{\vert k,C \ket} =:  d_{1}^{\vert k \ket} $, since 
they are evaluated for the unperturbed boundary states
which are  independent of $\mu$ and hence of $C$.

In the course of deriving the above formula, we have made several assumptions
following  \cite{Dorey:1999cj,Dorey:2005ak}:
the existence of  integrable boundary perturbations  of the HSG model
by operators with dimension $\Delta$, that of
the reflection factors associated with the identity operator, 
the invariance of the symmetry factor $C_{\vert \One \ket}$ under the
deformation, and the relation (\ref{muC}).
We will check that these are consistent with numerical
computations and the results in the conformal limit, which we discuss shortly.

\subsection{Identification of boundary conditions}

When evaluating the expansion
(\ref{Tcexpansion}), we need to identify the boundary conditions represented 
by the reflection factors (\ref{RkC}).
For this purpose, we first recall that
the Cardy boundary states are of the form
\eqb
\vert \alpha \ket = \sum_{\rho} \frac{S_{\alpha \rho}}{\sqrt{S_{0\rho}}} \vert \rho \ket \! \ket \comma
\eqe
where $ \vert \rho \ket \! \ket$ are the Ishibashi states.
In our case of the $\grp{SU}(n)_{2}/\grp{U}(1)^{n-1}$ coset CFT,
the modular S-matrix, $S_{\rho\rho'} $,  is given by the product of the S-matrix
for  $\grp{SU}(n)_{2}$  and the complex conjugate of the S-matrix
for  $\grp{U}(1)^{n-1}$ :
$S_{\rho\rho'} = S^{(2)}_{\brho\brho'} S^{\ast}_{\alg{u}(1)}$.
Here, $\brho := [\rho_{1}, ..., \rho_{n-1}]$ is the  Dynkin label of $\alg{su}(n)$.
Since we deal only with  primaries whose $ \alg{u}(1)$  weights are
zero, the $\alg{u}(1)$ part is trivial and hence we drop 
$S^{\ast}_{\alg{u}(1)}$ in the following.

In general, the index of the T-functions labels the representations
of the underlying symmetry. Thus, we infer that
$ R_j^{\vert k,C \ket}$ correspond to the $k$-th fundamental 
representation with the Dynkin label whose components are 
$(\brho_{k})_{j} = \delta_{j,k} $.
It then follows that
\eqb\label{gS}
g_{\vert \One \ket} 
  = \frac{S^{(2)}_{\bzero\bzero} }{\sqrt{S^{(2)}_{\bzero\bzero}}}
  \comma \quad 
  g_{\vert k \ket} 
  = \frac{S^{(2)}_{\brho_k\bzero} }{ \sqrt{ S^{(2)}_{\bzero\bzero} } } \period
\eqe
Here, the S-matrix for $\grp{SU}(n)_{2}$
is given by the formula
\cite{modularS}, 
\eqb\label{level2S}
 S^{(2)}_{\brho\bfm{\mu}}\Eqn{=}
(n+2)^{-(n-1)/2}\frac{i^{n(n-1)/2}}{\sqrt{n}}
\exp\left[
\frac{2\pi i}{n(n+2)}
\left(\sum_{j=1}^{n-1} j(\rho_j+1)\right)
\left(\sum_{j=1}^{n-1} j(\mu_j+1)\right)
\right],\nn\\
&&\times\det\left(
\exp\left[-\frac{2\pi i}{n+2}
\left(\sum_{j=a}^{n-1}(\rho_j+1)\right)
\left(\sum_{j=b}^{n-1}(\mu_j+1)\right)
\right]
\right)_{1\le a,b\le n} \period
\eqe
From (\ref{gS}) and (\ref{level2S}),
we  find  
the CFT limit of the T- and Y-functions:
\eqb\label{TYlim}
T_k \Eqn{\to}  c^{(0,0)}_{k} = \frac{S^{(2)}_{\brho_k\bzero} }{S^{(2)}_{\bzero\bzero} }
 =  \frac{\sin \frac{(k+1)\pi}{n+2}}{\sin \frac{\pi}{n+2}} 
\comma \nn \\
 Y_k 
\Eqn{\to}   \frac{S^{(2)}_{\brho_{k-1}\bzero} S^{(2)}_{\brho_{k+1}\bzero}  }{
 \bigl(S^{(2)}_{\bzero\bzero} \bigr)^{2}}
= \frac{\sin \frac{k\pi}{n+2} \sin \frac{(k+2)\pi}{n+2}}{\sin^2 \frac{\pi}{n+2}}
\period
\eqe
These agree with the result 
in  \cite{Alday:2010vh}, which 
supports our formula (\ref{Tcexpansion}) and identification 
of the boundary conditions.  
We note that the relation between the $g$-functions and the T-functions 
at the CFT point naturally explains 
the fact that the quantum dimensions (ratios
of the modular S-matrices) are the solutions
of the constant T-system, called the Q-system 
\cite{Kirillov:1987,Nahm:1992sx, Kuniba:2010ir}. 
From the relation between the T- and $g$-functions, 
this may hold for general TBA systems.

Since the bulk perturbing operator $\Phi$ is a linear combination
of the adjoint operators,  $g_{\vert \alpha \ket}^{\Phi}$ are similarly 
expressed by the elements of the modular S-matrix
with the Dynkin label of the adjoint representation $ \brho_{\rm adj}$:
\eqb\label{gPhi}
 g_{\vert \One \ket}^{\Phi} = G(\tilM_{i}) \frac{S^{(2)}_{
 \bzero\brho_{\rm adj} }}{\sqrt{S^{(2)}_{\bzero\brho_{\rm adj}}}}
 \comma \quad 
  g_{\vert k \ket}^{\Phi} = G(\tilM_{i}) \frac{S^{(2)}_{
  \brho_k\brho_{\rm adj} }}{\sqrt{S^{(2)}_{\bzero\brho_{\rm adj}}}} \period
\eqe
Thus, at the leading order, both of the expansions of 
the free energy and the Y-functions are given in terms of 
$G(\tilM_{j})$.
In addition, in order to obtain the  full expressions of the scattering 
amplitudes,  we need
the explicit form of $\Phi$, as well as 
the bulk coupling $\lambda$,  in terms of the TBA
masses $m_{k}$. These are discussed in section 6.

\section{Remainder function for the octagon}

As we have seen in section~2, the minimal surface with a null polygonal boundary in the AdS space 
is described by the Y-system or the TBA equations.
The Y-functions and the T-functions play important roles in this picture.
In the previous two sections, we have seen the relation between the $g$-functions and the T-functions
in the underlying integrable model.
This relation enables us to compute the high-temperature (small mass) expansions of the T-functions by using
the conformal perturbation technique.
Consequently, the remainder function is expanded around
the kinematic configurations associated with regular polygons.
Here we consider the first non-trivial example in $AdS_3$: the octagon.
In this case, the exact expression of the remainder function at strong coupling has been computed
by Alday and Maldacena \cite{Alday:2009yn}, and
one can learn much about the expansions of the T-functions and the $g$-functions.
The underlying integrable model corresponding to the octagon is the off-critical Ising model 
(with a complex mass).
The exact $g$-function for the off-critical Ising model was obtained in \cite{LeClair:1995uf, Chatterjee:1995be, Dorey:2004xk}.
Though the TBA system of the off-critical Ising model is trivial, the high-temperature expansions of
the free energy and of the T-function are still non-trivial.

As in \eqref{eq:Reven}, the remainder function is divided into several pieces:
\begin{align}
R_8=\frac{7\pi}{6}+A_{\rm free} + A_{ \rm periods} + A_{ \rm extra} + \Delta A_{ \rm BDS}.
\end{align}
These terms are given by
\begin{align}
A_{\rm free}&=\int_{-\infty}^\infty \frac{dt}{2\pi} l\cosh t\log\left(1+e^{-l \cosh t}\right), \label{eq:A_sinh}\\
A_{\rm periods}&=0, \\
A_{\rm extra}&=-\frac{l}{2}(\cos\phi \log\gamma_1^{\rm L}+\sin\phi\log\gamma_1^{\rm R}),\\
\Delta A_{\rm BDS}&=A_{\text{BDS-like}}-A_{\rm BDS}=-\frac{1}{2}\log(1+\chi^-)\log\( 1+\frac{1}{\chi^+}\), \label{eq:DelA_BDS}
\end{align}
where 
\begin{align}\label{chipm}
\chi^+=e^{l\sin\phi},\qquad \chi^-=e^{-l\cos\phi},
\end{align}
and $\gamma_1^{\rm L}, \gamma_1^{\rm R}$ are obtained 
from (\ref{gammaLR}) with 
\begin{align}
\log \gamma_1(\zeta=e^\theta)=\frac{1}{2\pi}\int_{-\infty}^\infty \frac{dt}{\cosh(t-\theta+i\phi)}
\log\left(1+e^{-l\cosh t}\right).
\label{eq:gamma1}
\end{align}
We note that \eqref{eq:gamma1} is a special case of \eqref{Teq2} 
for $k=1$ with $\nu_1=0$ and $Y_1(\theta)=e^{-l\cosh\theta}$,
in accord with \eqref{eq:gamma1T1}.
Our goal here is to expand the remainder function around $l=0$.
For the octagon, we can get the all-order expansion at arbitrary $\phi$.

The expansion of the free energy was studied in \cite{Klassen:1990dx}, and the result is the following,
\begin{align}
A_{\rm free}
&=\frac{\pi}{12}-\frac{l^2}{4\pi}\(\log\frac{1}{l}+\frac{1}{2}+\log\pi-\gamma_{\rm E}\)\notag \\
&\hspace{0.5cm}+\pi \sum_{k=1}^\infty \begin{pmatrix} \frac{1}{2} \\ k+1 \end{pmatrix}
\(1-\frac{1}{2^{2k+1}}\)\zeta(2k+1)\(\frac{l}{\pi}\)^{2k+2}, \label{eq:A_free-exp}
\end{align}
where $\gamma_{\rm E}$ is the Euler constant.

In order to expand $A_{\rm extra}$, we introduce the following function,
\begin{align}
F(l,\varphi)=\frac{1}{2\pi}\int_{-\infty}^\infty \frac{dt}{\cosh(t+i\varphi)}\log(1+e^{-l \cosh t}) \qquad
\( |\varphi|<\frac{\pi}{2}\),
\label{eq:F(l,phi)}
\end{align}
where $\gamma_1^{\rm L}$ and $\gamma_1^{\rm R}$ are related to this function as
\begin{align}
\log \gamma_1^{\rm L}=F(l,\phi),\qquad
\log \gamma_1^{\rm R}=F\(l,\phi-\frac{\pi}{2}\).
\end{align}
Note that this function is related to the exact $g$-function for the off-critical Ising model considered in \cite{Dorey:2004xk}.
In \cite{Dorey:2004xk}, the exact $g$-function was discussed in two special cases,
which essentially reduce to $\varphi=0$ in our case.
$\varphi \ne 0$ corresponds to the case that the boundary magnetic field is turned on in the off-critical Ising model.
Here we obtain the small $l$ expansion of the $g$-function for general values of $\varphi$ with $|\varphi|<\pi/2$.
As we will discuss in Appendix A, this function has the expansion
\begin{align}
&F(l,\varphi)\nn\\
&=\frac{1}{2}\log(1+e^{-l \sin \varphi})
-\frac{l}{2\pi}\biggl[ \cos\varphi\(\log\frac{1}{l}+1+\log\pi-\gamma_{\rm E}\)+\(\varphi-\frac{\pi}{2}\) \sin \varphi\biggr]\notag \\
&+\sum_{k=1}^\infty \begin{pmatrix} -\frac{1}{2} \\ k \end{pmatrix}
\frac{1}{2k+1}\(1-\frac{1}{2^{2k+1}}\)\zeta(2k+1)\cos\varphi\;{}_2F_1(-k,1;\frac{1}{2}-k;\sin^2\varphi)\(\frac{l}{\pi}\)^{2k+1}\!.
\label{eq:F-exp1}
\end{align}
Using this expansion, we find the small $l$ expansion of $A_{\rm extra}$,
\begin{align}
A_{\rm extra}
&=-\frac{1}{4}l\cos\phi\log(1+e^{-l\sin\phi})-\frac{1}{4}l\sin\phi\log(1+e^{-l\cos\phi}) \notag \\
&\hspace{0.5cm}+\frac{l^2}{4\pi}\(\log\frac{1}{l}+1+\log\pi-\gamma_{\rm E}-\frac{\pi}{2}\cos\phi\sin\phi\) \notag \\
&\hspace{0.5cm}-\pi \sum_{k=1}^\infty \begin{pmatrix} \frac{1}{2} \\ k+1 \end{pmatrix}
\(1-\frac{1}{2^{2k+1}}\)\zeta(2k+1)\frac{k+1}{2k+1}f_k(\phi)\(\frac{l}{\pi}\)^{2k+2}, \label{eq:A_extra-exp}
\end{align}
where the function $f_k(\phi)$ is expressed in terms of the hypergeometric functions,
\begin{align}
f_k(\phi)\equiv \cos^2\phi\; {}_2F_1(-k,1;\frac{1}{2}-k;\sin^2\phi)+\sin^2\phi\; {}_2F_1(-k,1;\frac{1}{2}-k;\cos^2\phi).
\end{align}
Combining \eqref{eq:DelA_BDS}, \eqref{eq:A_free-exp} and \eqref{eq:A_extra-exp}, 
we obtain the expression,
\begin{align}
R_8&=\frac{5\pi}{4}-\frac{1}{2}\log\(2\cosh \frac{l\cos\phi}{2}\)\log\(2\cosh \frac{l\sin\phi}{2}\)+\frac{l^2}{8\pi}\notag \\
&\hspace{0.5cm}+\pi \sum_{k=1}^\infty \begin{pmatrix} \frac{1}{2} \\ k+1 \end{pmatrix}
\(1-\frac{1}{2^{2k+1}}\)\zeta(2k+1)\(1-\frac{k+1}{2k+1}f_k(\phi)\)\(\frac{l}{\pi}\)^{2k+2}.
\label{eq:R8-exp}
\end{align}
Note that the non-analytic terms in \eqref{eq:A_free-exp} and \eqref{eq:A_extra-exp} cancel each other out. 
The expression \eqref{eq:R8-exp} is convergent for $|l|<\pi$. 
One can immediately confirm that the remainder function is expanded in $l^2$:
\begin{align}\label{R8}
R_8=\sum_{k=0}^\infty R_8^{(2k)}(\phi)l^{2k},
\end{align}
where the first four coefficients are given by
\begin{align}
R_8^{(0)}(\phi)&=\frac{5\pi}{4}-\frac{\log^2 2}{2}, \\
R_8^{(2)}(\phi)&=\frac{1}{8\pi}-\frac{\log 2}{16}, \\
R_8^{(4)}(\phi)&=\frac{2\log2-1}{1024}+\(\frac{2\log2+3}{3072}-\frac{7\zeta(3)}{192\pi^3}\)\cos4\phi,\\
R_8^{(6)}(\phi)&=\frac{-8\log2+3}{73728}-\(\frac{8\log2+5}{122880}-\frac{31\zeta(5)}{1280\pi^5}\)\cos4\phi.
\end{align}
We note that these maintain the symmetries $\phi \to - \phi$ and
$\phi \to \phi+\frac{\pi}{2}$, which are due to the space-time parity 
and cyclicity \cite{Alday:2009yn}.
The expansion \eqref{eq:A_extra-exp} was derived for $0<\phi<\pi/2$,
but is valid for arbitrary $\phi$ due to these symmetries.
In section~7, we will compare these results with the remainder function at two loops.

\section{High-temperature expansion for the decagon}

In the previous section, we have considered the remainder function for the octagon,
and have obtained its all-order expansion with respect to the mass scale parameter $l$.
The crucial point there is that the TBA system for the octagon is trivial.
For the general $2\tilde{n}$-gon ($\tilde{n}\geq 5$), however, 
analytic solutions of the TBA equations have not been known yet.
In this section, we consider the second simplest case
$\tilde{n}=5$: the decagon,
and see how to compute the high-temperature expansion of its remainder function.
Here, the underlying integrable theory is the homogeneous sine-Gordon model associated with the $ \grp{SU}(3)_2/ \grp{U}(1)^2$ coset CFT.

The remainder function for the decagon is divided into the following parts,
\begin{align}
R_{10} = \frac{7}{4} \pi + A_{\rm periods} + A_{\rm free} + \Delta A_{\rm BDS}.
\end{align}
The period part is given by
\begin{align}
A_{\rm periods} = -\frac{1}{4} (m_{1} \bar{m}_{2}+ m_{2} \bar{m}_{1})=-\frac{1}{2}\tilde{M}_1 \tilde{M}_2 l^2\cos(\varphi_1-\varphi_2),
\end{align}
where $m_j=M_j L e^{i\varphi_j}$, $\tilde{M}_j=M_j/M$ and $l=ML$.
The free energy part is written as
\begin{align}
A_{\rm free}=\sum_{j=1}^2 \int_{-\infty}^\infty \frac{d\theta}{2\pi} M_j L \cosh \theta
\log(1+\tilde{Y}_j(\theta)), 
\end{align}
where  $\tilde{Y}_j(\theta)$ is defined in (\ref{Ytilde}).
From \eqref{DeltaBDS1}, the part $\Delta A_{\rm BDS}$ is given by
\begin{align}\label{DBDS10}
\Delta A_{ \rm BDS}  
= \frac{1}{4} \sum_{i,j = 1}^{5} \log\frac{ c_{i,j}^{+} }{ c_{i,j+1}^{+} } 
\log \frac{ c_{i-1,j}^{-} }{c_{i,j}^{-} }.
\end{align}
For the decagon, there are four independent cross-ratios.
This is consistent with the fact that the TBA system has two independent complex parameters $m_j$ ($j=1,2$).
The cross-ratios $c_{13}^\pm$ and $c_{14}^\pm$ are related to the Y-functions,
\begin{align}
c_{13}^-&=Y_1(0),\qquad
c_{13}^+=Y_1\(-\frac{\pi i}{2}\), \label{eq:c13}\\
c_{14}^+&=Y_2(0),\qquad
c_{14}^-=Y_2\(\frac{\pi i}{2}\). \label{eq:c14}
\end{align}
The other cross-ratios $c^{\pm}_{24}$, $c^{\pm}_{25}$ and $c^{\pm}_{35}$ are expressed by $c^{\pm}_{13}$
and $c^{\pm}_{14}$, 
\begin{align}\label{c-others}
c^{\pm}_{24} = \frac{1+c^{\pm}_{14}}{c^{\pm}_{13}}, \quad 
c^{\pm}_{35} = \frac{1+c^{\pm}_{13}}{c^{\pm}_{14}}, \quad 
c^{\pm}_{25} =  \( 1+ \frac{1}{c^{\pm}_{13}}\) 
 \( 1+ \frac{1}{c^{\pm}_{14}}\) -1.
\end{align}
Our goal is to find the small $l$ behaviors of the remainder function.

\subsection{Case with real masses}

Let us start by restricting our attention to the case that two masses are real: $\varphi_1=\varphi_2=0$.
In this case, we can directly use the results of the CPT in section~4.
The period term is
\begin{align}
A_{\rm periods}=-\frac{1}{2}\tilde{M}_1 \tilde{M}_2 l^2.
\label{eq:periods-real}
\end{align}
Let us consider the free energy part.
Since the central charge of the $\grp{SU}(3)_2/ \grp{U}(1)^2$
coset CFT is
$c_3=6/5$, the free energy goes to $\pi/5$ in the limit $(m_1,m_2) \to (0,0)$.
The perturbing operator has the dimension $\Delta=\bar{\Delta}=3/5$.
From \eqref{eq:Bnodd}, the bulk term is given by
\begin{align}
f_3^{\rm bulk}=\frac{1}{2}\tilde{M}_1\tilde{M}_2 l^2.
\end{align}
Thus the free energy has the following expansion,
\begin{align}
A_{\rm free}=\frac{\pi}{5}+\frac{1}{2}\tilde{M}_1\tilde{M}_2 l^2+\sum_{k=2}^\infty f_3^{(k)}l^{4k/5} ,
\label{eq:free-real}
\end{align}
where $f_3^{(k)}$ is computed by the CPT (see \eqref{CPTF}).
In particular, $f_3^{(2)}$ is read from (\ref{fn2}):
\eqb
f_{3}^{(2)} = \frac{\pi}{6}  \kappa_{3}^{2}  G^{2}(\tilde{M}_1,\tilde{M}_2)  C_{3}^{(2)}
\comma \quad C_{3}^{(2)} = 3 (2\pi)^{\frac{4}{5}}
\gamma^{2}\Bigl( \frac{3}{5}\Bigr)  \gamma\Bigl( -\frac{1}{5}\Bigr) ,
\label{eq:f32}
\eqe
where $\gamma(x)=\Gamma(x)/\Gamma(1-x)$ and
\begin{align}
G(\tilde{M}_1,\tilde{M}_2)=\sum_{i,j=1}^2 \tilde{M}_i^{2/5}F_{ij}\tilde{M}_j^{2/5}.
\label{eq:G-1}
\end{align}
We need to determine the symmetric matrix $F$,
which has two independent components for the decagon.
In principle, it should be possible to fix them by considering the quantum theory
of the SU(3)$_2/$U(1)$^2$ HSG model.
However, we take a different strategy here. 
Fortunately, we can completely fix $F_{ij}$ by the following consideration in the case of the decagon.

Let us first take the limit $(M_1,M_2) \to (M,0)$.
In this limit, the TBA equations for the decagon reduce to those for the (RSOS)$_3$ scattering theory.
Therefore the correction \eqref{eq:f32} should be equal to that for the (RSOS)$_3$ scattering theory,%
\footnote{Since the (RSOS)$_3$ model has the central charge $c=7/10$, the free energy goes to $7\pi/60$ in the limit $ML\to 0$.
The difference of the constant terms of the free energies in two theories 
comes from the fact that the contribution of particle 2 is absent
in the (RSOS)$_3$ scattering theory. One can numerically check that in the homogeneous sine-Gordon model,
the contributions of particles 1 and 2 go to $7\pi/60$ and $\pi/12$, respectively in the limit $ML \to 0$ 
with $\tilde{M}_2 \ll \tilde{M}_1=1$, 
and the sum of two-particle contributions gives the correct value $\pi/5$.
Furthermore all the high-temperature corrections of particle 2 should vanish if we take the limit $M_2 \to 0$.}
and we obtain
\begin{align}
F_{11}\kappa_3=\kappa_3^{{\rm RSOS}} ,
\label{eq:kappa-rel}
\end{align}
where $\kappa_3^{{\rm RSOS}}$ is given by \eqref{eq:kappaRSOS}.

\begin{figure}[tbp]
\begin{center}
\begin{tabular}{cc}
\hspace{-3mm}
\resizebox{75mm}{!}{\includegraphics{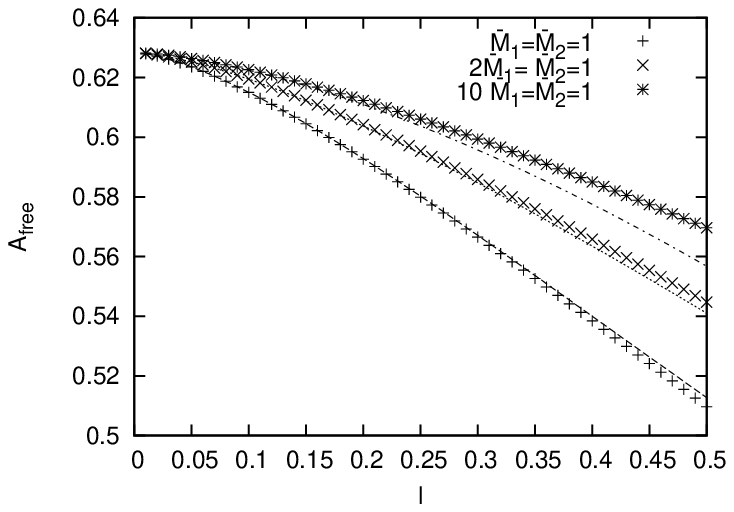}}
\hspace{-4mm}
&
\resizebox{75mm}{!}{\includegraphics{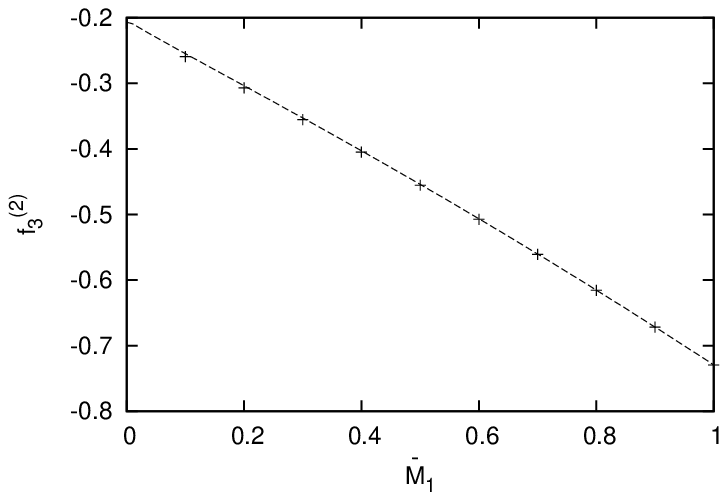}}
\hspace{-5mm}
\\
(a) & (b)
\end{tabular}

\end{center}
\caption{(a) The scale parameter $l$-dependence of the free energy with
fixed $\tilde{M}_1/\tilde{M}_2=1,1/2$ and $1/10$.
Dashed lines
represent the curve
$\frac{\pi}{5}+\frac{1}{2}\tilde{M}_1\tilde{M}_2l^2+f^{(2)}_3l^{8/5}$.
Deviation from the analytic formula at $l=0.5$ for
$\tilde{M}_1/\tilde{M}_2=1/10$
comes from the next order correction $O(l^{12/5})$, which is estimated from
the numerical fit as  $0.08 l^{12/5}\approx 0.015$.
(b) $\tilde{M}_1$-dependence of the coefficient of $l^{8/5}$ of the free
energy for $\tilde{M}_2=1$. Dashed line corresponds to the curve
$f^{(2)}_3$ given in \eqref{eq:f32-2}.}

\label{fig1}
\end{figure}

Let us next consider the case: $M_1=M_2=M$.
In this case, the TBA equations are regarded as those associated with the tadpole Dynkin diagram $T_1$.
The corresponding integrable model is the non-unitary $(\grp{SU}(2)_{-1/2} \times \grp{SU}(2)_1)/\grp{SU}(2)_{1/2}$ 
coset model perturbed by the primary field $\phi_{1,1,3}$ with dimension $\Delta=\bar{\Delta}=3/5$ (see subsection~3.2). 
We discuss the perturbation of the above coset model in Appendix~B.
Using the result \eqref{eq:free-M1=M2}, we find the constraint
\begin{align}
\frac{2\pi}{3}\kappa_3^2(F_{11}+F_{12})^2C_3^{(2)}
=\frac{1}{8}\(\frac{\pi}{4}\)^{1/5}\gamma\(\frac{1}{4}\)^{8/5}\gamma\(-\frac{1}{5}\)
\gamma\(\frac{3}{5}\)\gamma\(\frac{4}{5}\).
\label{eq:F11+F12}
\end{align}
Combining \eqref{eq:kappa-rel} and \eqref{eq:F11+F12}, we obtain
\begin{align}
1+\frac{F_{12}}{F_{11}}=\frac{1}{2}\(\frac{3}{\pi^2}\)^{1/5}\gamma\(\frac{1}{4}\)^{4/5}.
\end{align}
From these two considerations, the order $l^{8/5}$ correction of the free energy must have the following form
\begin{align}
f_3^{(2)}=f_{({\rm RSOS})_3}^{(2)}
F_{11}^{-2}G^2(\tilde{M}_{1}, \tilde{M}_{2})
\label{eq:f32-2}
\end{align}
where
\begin{align}
f_{({\rm RSOS})_3}^{(2)}&=\frac{\pi}{6}(\kappa_3^{\rm RSOS})^2 C_3^{(2)}
=\frac{\pi}{8\cdot 6^{2/5}} \gamma\(-\frac{1}{5}\)\gamma\( \frac{3}{5}\) \gamma\(\frac{4}{5}\),
\end{align}
and
\begin{align}
F_{11}^{-1}G(\tilde{M}_{1}, \tilde{M}_{2})&=\tilde{M}_1^{4/5}+\tilde{M}_2^{4/5}-B \tilde{M}_1^{2/5}\tilde{M}_2^{2/5},
\label{eq:G-2}\\
B&= - \frac{2F_{12}}{F_{11}}
=2-\(\frac{3}{\pi^2}\)^{1/5}\gamma\(\frac{1}{4}\)^{4/5}.
\end{align}
As seen in section~3, the matrix $F$ deviates from the inverse of the Cartan matrix.
We have confirmed that \eqref{eq:f32-2} is in good agreement with
the numerical results for arbitrary
$\tilde{M}_1$ and $\tilde{M}_2$ (Fig.~\ref{fig1}). 

Now we proceed to the expansion of $\Delta A_{\rm BDS}$.
In order to know the small $l$ behavior of $\Delta A_{\rm BDS}$, we need
the high-temperature expansions of the Y-functions.
Since the Y-functions are related to the T-functions, we can use the results in section~4.
In the decagon case, the relations between $Y_j$ and $T_j$ are as follows,
\begin{align}
Y_1(\theta)=T_2(\theta),\quad Y_2(\theta)=T_1(\theta).
\end{align}
As in (4.19) and (4.27), the high-temperature expansions of the T-functions are computed
from the data of the $g$-functions at the CFT point.
Using the formula (4.30) of the modular S-matrix, one obtains
\begin{align}
\frac{g_{\vert 1 \ket}}{g_{\vert \One \ket}}&=\frac{g_{\vert 2 \ket}}{g_{\vert \One \ket}}
=2\cos\(\frac{\pi}{5}\),\quad
\frac{g_{\vert \One \ket}^\Phi}{g_{\vert \One \ket}}
=G(\tilde{M}_1,\tilde{M}_2)\left[ 2\cos\(\frac{\pi}{5}\) \right]^{1/2},\\
\frac{g_{\vert 1 \ket}^\Phi}{g_{\vert 1 \ket}}
&=\frac{g_{\vert 2 \ket}^\Phi}{g_{\vert 2 \ket}}
=-G(\tilde{M}_1,\tilde{M}_2)\left[ 2\cos\(\frac{\pi}{5}\) \right]^{-3/2},
\end{align}
Therefore from \eqref{Tcexpansion}, we find
\begin{align}
Y_j(\theta)=Y^{(0)}+Y^{(2)}(\tilde{M}_1,\tilde{M}_2)l^{4/5}\cosh\(\frac{4\theta}{5}\)+{\cal O}(l^{6/5}),
\label{eq:Y-exp-real}
\end{align}
where
\begin{align}
Y^{(0)}=2\cos\(\frac{\pi}{5}\),
\end{align}
and
\begin{align}
Y^{(2)}(\tilde{M}_1,\tilde{M}_2)
&=y_{({\rm RSOS})_3}^{(2)}
F_{11}^{-1}G(\tilde{M}_1,\tilde{M}_2)
\label{eq:exp_Y}\\
y_{({\rm RSOS})_3}^{(2)}&=\frac{1}{4\cdot6^{1/5}}\Gamma\(-\frac{1}{5}\)
\left[ 10\cos\(\frac{\pi}{5}\) \gamma\(\frac{3}{5}\)\gamma\(\frac{4}{5}\) \right]^{1/2}.
\label{eq:y_RSOS}
\end{align}
From \eqref{eq:c13} and \eqref{eq:c14}, the cross-ratios are expanded as
\begin{align}
c_{13}^{-} = c_{14}^{+} &= Y^{(0)} + Y^{(2)} l^{4/5}+ {\cal O}(l^{6/5}), \label{eq:c_13^-}\\
c_{13}^{+} = c_{14}^{-} &= Y^{(0)} + Y^{(2)} l^{4/5}\cos\( \frac{2}{5}\pi\) + {\cal O}(l^{6/5}) .
\label{eq:c_13^+}
\end{align}
As shown in Appendix D, the expansion of 
the $\Delta A_{\rm BDS}$  is largely constrained by the Y-system, 
the structure of the conformal perturbation, as well as 
the symmetries associated with the space-time parity and cyclicity,
under which $\Delta A_{\rm BDS}$ is invariant.
Consequently, 
it turns out that the terms of $\calO(l^{4/5})$ are enough to give
the expansion of $\Delta A_{\rm BDS}$ up to $\calO(l^{12/5})$. We then obtain
the high-temperature expansion of $\Delta A_{\rm BDS}$,
\begin{align}
\Delta A_{\rm BDS} =-\frac{5}{2}\log^2 \( 2\cos\(\frac{\pi}{5}\) \)
 +  B_{2} ( Y^{(2)} )^{2} l^{8/5} + {\cal O}(l^{12/5}) ,
 \label{eq:dA_BDS-real}
\end{align}
where $B_{2}$ 
is given by
\begin{align}
B_2 = 20 \cos^4\( \frac{2\pi}{5} \)
 \( 1- \frac{1}{ \sqrt{5} }  \log \( 2 \cos \(\frac{\pi}{5} \) \) \).
\label{eq:B_2}
\end{align}
Note that the first order term ${\cal O}(l^{4/5})$ vanishes.
In summary, from \eqref{eq:periods-real}, \eqref{eq:free-real}, \eqref{eq:f32-2} and \eqref{eq:dA_BDS-real},
the remainder function with the real masses has the following expansion,
\begin{align}
R_{10}=R_{10}^{(0)}+R_{10}^{(4)}l^{8/5}+{\cal O}(l^{12/5}),
\end{align}
where
\begin{align}
R_{10}^{(0)}&=\frac{39}{20}\pi-\frac{5}{2}\log^2\( 2\cos\(\frac{\pi}{5}\)\),\\
R_{10}^{(4)}&
=\(-\frac{1}{5}\tan\(\frac{\pi}{5}\)+B_2\)Y^{(2)}(\tilde{M}_1,\tilde{M}_2)^2.
\end{align}
Note that $A_{\rm periods}$ is canceled by the bulk term in the free energy, and the remainder function
is expanded in $l^{2/5}$.
We also comment that the ratio $f^{(2)}_{\rm (RSOS)_{3}}/(y^{(2)}_{\rm (RSOS)_{3}})^{2}$ 
interestingly becomes very simple,
\begin{align}
\frac{f^{(2)}_{\rm (RSOS)_{3}}}{(y^{(2)}_{\rm (RSOS)_{3}})^{2}}=-\frac{1}{5}\tan\(\frac{\pi}{5}\).
\end{align}

\subsection{Case with complex masses}

So far, we have focused on the case that two masses are real.
We would  now like to discuss the  general situation 
where two masses are complex.
The phase of the complex mass corresponds to the purely imaginary 
resonance parameter. As discussed in section 3, 
it is not clear if this case can be treated 
within the framework of the conformal perturbation of the HSG model.
However, one can expect that the expansion is analytic in the mass parameters
\cite{Dorey:1997rb}, and the expansion for the real masses can be extended to 
that for the complex masses by continuing the mass parameters.
This is also expected from the point of view of the TBA equations.
In fact, we will see that the results obtained in this way  are in agreement 
with numerical computations.
Furthermore, one can arrive at the same conclusion for some relevant 
quantities by considering  the chiral limit of the TBA system,
which is discussed in detail in Appendix C.

The way to incorporate the phase is determined by the
the following facts: the resonance parameters in the TBA equations
are understood as due to the rescaling of the mass parameters, 
the free energy, by definition, should depend only on the difference 
of the phases $\varphi_{12}=\varphi_{1} -\varphi_{2}$, and 
 $\bfm{\lambda}, \bar{\bfm{\lambda}}$ are of the form
(\ref{lambdapm}) semi-classically.
We thus  make a replacement, in the complex mass case,
\begin{align}
\boldsymbol{\lambda} &\to \sum_j (\tilde{M}_j e^{i\varphi_j})^{1-(\Delta+\bar{\Delta})/2} \boldsymbol{\hat{\lambda}}_j,\\
\bar{\boldsymbol{\lambda}} &\to \sum_j (\tilde{M}_j e^{-i\varphi_j})^{1-(\Delta+\bar{\Delta})/2} \boldsymbol{\hat{\lambda}}_j.
\end{align}
Then the two-point function of the perturbing operator becomes
\begin{align}
\bigl\langle \Phi_{\boldsymbol{\lambda},\bar{\boldsymbol{\lambda}}}(z)
\Phi_{\boldsymbol{\lambda},\bar{\boldsymbol{\lambda}}}(0) \bigr\rangle 
=\frac{\bigl|G(\tilde{M}_1e^{i\varphi_1},\tilde{M}_2e^{i\varphi_2})\bigr|^2}{|z|^{12/5}},
\end{align}
where $G$ is given by \eqref{eq:G-2}.

Let us consider the free energy.
As mentioned above,
the free energy must be a function of $\varphi_{12}$.
This suggests that the bulk term is modified as
\begin{align}
f_3^{\rm bulk} \to \frac{1}{4}l^2 \sum_{i,j=1}^2 (\tilde{M}_ie^{i\varphi_i})(I^{-1})_{ij} 
(\tilde{M}_j e^{-i\varphi_j})
=\frac{1}{2}\tilde{M}_1\tilde{M}_2 l^2\cos \varphi_{12} .
\end{align}
Taking into account these modifications, we find that the expansion of the free energy is given by
\begin{align}
A_{\rm free}=\frac{\pi}{5}+\frac{1}{2}\tilde{M}_1\tilde{M}_2l^2\cos \varphi_{12}+f_{({\rm RSOS})_3}^{(2)}
F_{11}^{-2}\bigl|G(\tilde{M}_1e^{i\varphi_1},\tilde{M}_2e^{i\varphi_2})\bigr|^2l^{8/5}+{\cal O}(l^{12/5}) .
\end{align}

The relation between the $g$- and T-functions in section 4
is not  applied to the case of complex masses. However, 
by similarly complexifying the mass parameters, we obtain 
the expansion of the Y-functions,
\begin{align}
Y_j(\theta)=2\cos\(\frac{\pi}{5}\)+
&\frac{1}{2}\Bigl(Y^{(2)}(\tilde{M}_1e^{-i\varphi_1},\tilde{M}_2e^{-i\varphi_2})e^{4\theta/5}\nn \\
&+Y^{(2)}(\tilde{M}_1e^{i\varphi_1},\tilde{M}_2e^{i\varphi_2})e^{-4\theta/5}\Bigr)
l^{4/5}+{\cal O}(l^{6/5}),
\label{eq:exp-Y}
\end{align}
where $Y^{(2)}$ is given by \eqref{eq:exp_Y}.
We have checked that this formula agrees with
the numerical results (Fig.~\ref{fig2}).
The expansion of the Y-functions for the complex masses 
is also discussed  in Appendix~C  from the  chiral limit of the TBA system.
The space-time cross-ratios are again obtained by using the relations \eqref{eq:c13} and \eqref{eq:c14}.
\begin{figure}[tbp]
\begin{center}
\begin{tabular}{cc}
\hspace{-3mm}
\resizebox{75mm}{!}{\includegraphics{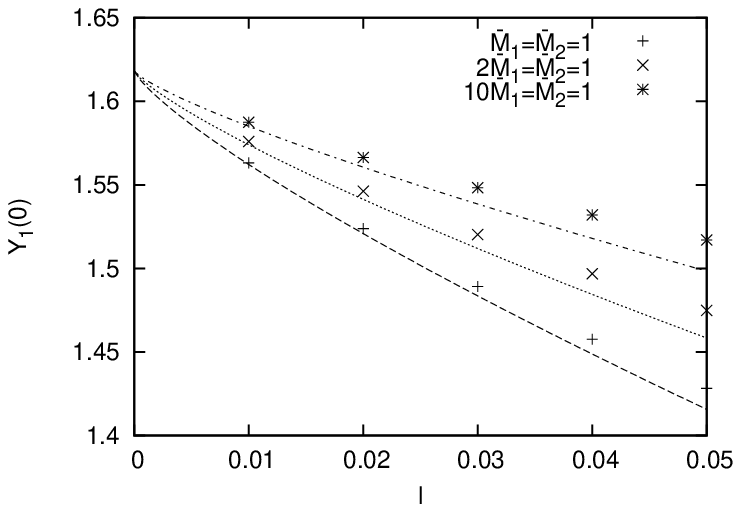}}
&
\hspace{-4mm}
\resizebox{75mm}{!}{\includegraphics{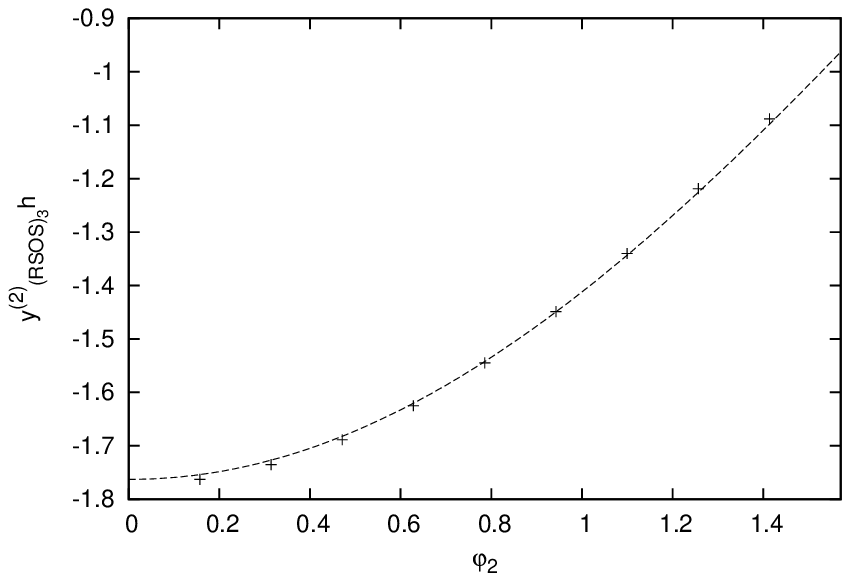}}
\hspace{-5mm}
\\
(a) & (b)
\end{tabular}
\end{center}
\caption{(a) The $l$-dependence of the Y-function $Y_1(0)$
for fixed $\tilde{M}_1/\tilde{M}_2=1,1/2$ and $1/10$
at $\varphi_1=\varphi_2=\pi/20$.
Dashed lines
represent the curve \eqref{eq:exp-Y} at $\theta=0$ up to the order $l^{4/5}$.
Deviation from the analytic formula comes from the next order correction
$O(l^{6/5})$, which can be estimated from the numerical fit as $0.15
l^{6/5}\approx 0.004$ ($l=0.05$) for $\tilde{M}_1=\tilde{M}_2=1$.
(b) Plots of the coefficient of $l^{4/5}$ in $Y_1(0)$ for 
$\varphi_1=\pi/20$
and various $\varphi_2$ at $2\tilde{M}_1=\tilde{M}_2=1$.
Dashed line
represents the curve
$\frac{1}{2}\bigl(Y^{(2)}(\tilde{M}_1e^{-i\varphi_1},\tilde{M}_2e^{-i\varphi_2})
+Y^{(2)}(\tilde{M}_1e^{i\varphi_1},\tilde{M}_2e^{i\varphi_2})\bigr)
=: y^{(2)}_{\rm  (RSOS)_{3}} h(\tilde{M}_{j},\varphi_{j})$.}
\label{fig2}
\end{figure}
In addition, using \eqref{eq:exp-Y}, 
the expansion of $\Delta A_{\rm BDS}$ is given 
as in the case of the real masses by
\begin{align}
\Delta A_{\rm BDS}=-\frac{5}{2}\log^2 \( 2\cos\(\frac{\pi}{5}\) \)
+B_2\bigl|Y^{(2)}(\tilde{M}_1e^{i\varphi_1},\tilde{M}_2 e^{i\varphi_2})\bigr|^2
l^{8/5}+{\cal O}(l^{12/5}).
\end{align}

\begin{figure}[tbp]
\begin{center}
\begin{tabular}{cc}
\hspace{-3mm}
\resizebox{75mm}{!}{\includegraphics{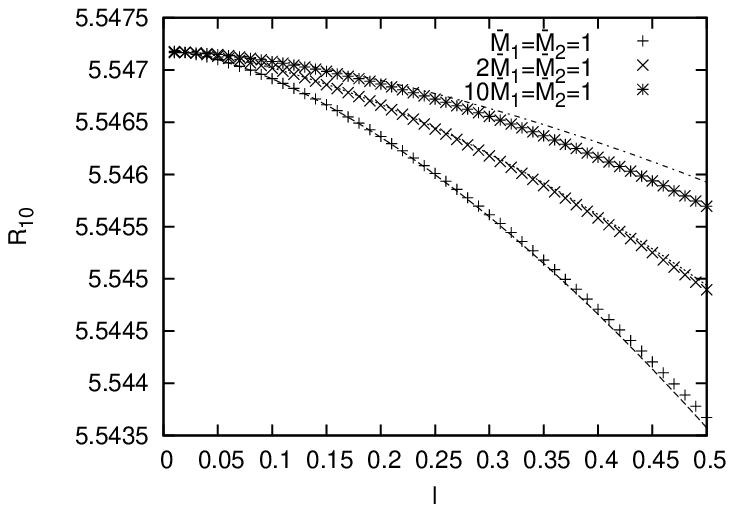}}
&
\hspace{-4mm}
\resizebox{75mm}{!}{\includegraphics{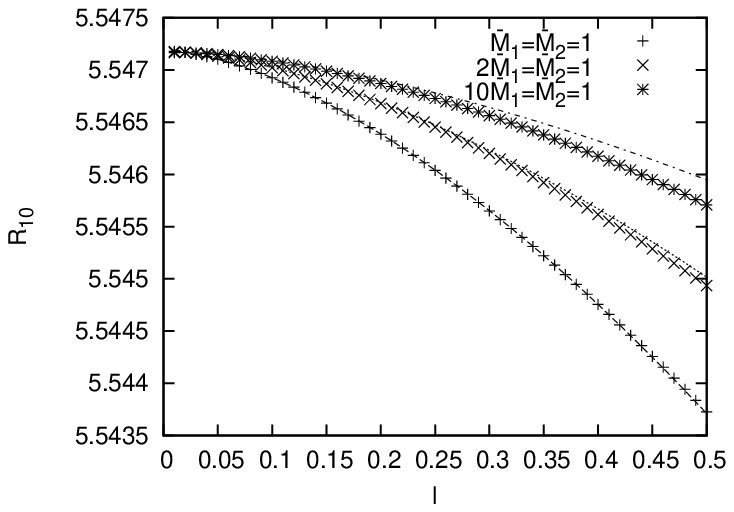}}
\hspace{-5mm}
\\
(a) & (b)
\end{tabular}
\end{center}
\caption{The $l$-dependence of the remainder function with (a)
equal phase $\varphi_1=\varphi_2={\pi/20}$
(b) different phase $\varphi_1=\pi/20$, $\varphi_2=\pi/5$. Dashed lines
represent the curve $R_{10}^{(0)}+R_{10}^{(4)}l^{8/5}$.}
\label{fig3}
\end{figure}

Collecting all the above results, we finally find that the remainder function with the complex masses 
has the expansion,
\begin{align}
R_{10}=R_{10}^{(0)}+R_{10}^{(4)}l^{8/5}+{\cal O}(l^{12/5}),
\end{align}
where
\begin{align}
R_{10}^{(0)}&=\frac{39}{20}\pi-\frac{5}{2}\log^2\( 2\cos\(\frac{\pi}{5}\)\),\\
R_{10}^{(4)}&=\(-\frac{1}{5}\tan\(\frac{\pi}{5}\)+B_2\)
\bigl|Y^{(2)}(\tilde{M}_1e^{i\varphi_1},\tilde{M}_2 e^{i\varphi_2})\bigr|^2.
\end{align}
$Y^{(2)}$ and $B_2$ are defined by 
\eqref{eq:exp_Y} and \eqref{eq:B_2}, respectively.
Note that the $A_{\rm periods}$ is canceled by the bulk term of the free energy again.
We have confirmed that this formula for $R_{10}$
is  in good agreement with the
numerical results for various values of 
$\tilde{M}_{1,2}$ and $\varphi_{1,2}$
(Fig.~\ref{fig3}).

At the end of this section, we comment on the relation between cross-ratios and the parameters 
in the TBA system.
In order to express the remainder function as a function of the cross-ratios,
one has to invert the relations  \eqref{eq:c13} and \eqref{eq:c14}. 
This is complicated for general complex masses. However, when 
the  phases of $m_{j}$ are equal, i.e., $\varphi_{j} = \varphi$,
$Y_{j}(\theta)$ are obtained from those for the real masses
by the shift $\theta \to \theta - \varphi$.
One then simply has
\eqb
 \frac{4}{5} \varphi \Eqn{=} 
 \tan^{-1} \( \cot \(\frac{2}{5}\pi \) \frac{ \delta c_{14}^{-}
  - \delta c_{13}^{+} }{ \delta c_{14}^{-} + \delta c_{13}^{+} }\) \comma  \quad 
 Y^{(2)} l^{4/5} = \frac{\delta c_{13}^{+}}{
  \cos\bigl( \frac{2}{5}(\pi + 2\varphi )\bigr) } 
  \comma
\eqe
where $\delta c_{jk}^{\pm} := c_{jk}^{\pm} -Y^{(0)}$.
Of course, one has to keep in mind that these expressions are 
valid for small $l$.
This corresponds to focusing on the kinematics near $c_{13}^{+}
=c_{13}^{-}=c_{14}^{+}=c_{14}^{-}$
in the space of the cross-ratios (or equivalently near the regular decagon).

%

\section{Comparison with  two-loop results}

 Wilson loops with light-like edges are  dual
to gluon scattering  amplitudes \cite{Dr}. For the kinematic configurations
corresponding to the $AdS_{3}$ octagon, the analytic
expression of the  2-loop remainder function for the Wilson loop
is given in  \cite{DelDuca:2010zp}. 
The analytic expression 
for the case of $AdS_{3}$ $2n$-gon has also been  written down 
\cite{Heslop:2010kq,Gaiotto:2010fk}. In this section, we compare 
our strong coupling results with those at  two loops
as expansions around the kinematic configurations associated 
with regular polygons.

\subsection{Octagon}

In the case of the octagon, the  remainder function at   two loops is\footnote{
The overall coupling dependence is suppressed.
} 
\eqb
R_{8}^{\rm 2\mbox{-}loop}
=-\frac{\pi^4}{18}-\frac{1}{2}\log(1+\chi^+)\log\(1+\frac{1}{\chi^+}\)
\log(1+\chi^-)\log\(1+\frac{1}{\chi^-}\) \comma
\eqe 
where the cross-ratios $\chi^{\pm}$ are given in (\ref{chipm}).
Similarly to the strong coupling case, this is expanded by using (\ref{logcosh}) 
as 
\eqb\label{Oct2L}
R_{8}^{\rm 2\mbox{-}loop} 
= \sum_{k=0}^\infty R_{8}^{{\rm 2\mbox{-}loop}\,   (2k)} (\phi) \, l^{2k}
\comma
\eqe
where the first few coefficients are
\eqb
R_{8}^{\rm 2\mbox{-}loop \,(0)}(\phi) \Eqn{=} -\frac{\pi^4}{18}-\frac{\log^4 2}{2}, \nn \\
R_{8}^{\rm 2\mbox{-}loop \, (2)}(\phi) \Eqn{=} -\frac{\log^2 2(\log 2-1)}{8}, \nn \\
R_{8}^{\rm 2\mbox{-}loop \, (4)}(\phi) \Eqn{=}
\frac{2\log^3 2-5\log^2 2+4\log 2-2}{512}   \\
&&  \qquad + \ \frac{2\log^3 2+3\log^2 2-12\log 2+6}{1536}\cos4\phi .\nn 
\eqe
One can check that $ R_{8}^{\rm 2\mbox{-}loop \,(0)}$ 
agrees with $R_{8}^{\rm 2\mbox{-}loop}$ 
for the regular octagon \cite{DelDuca:2010zp},
and that the coefficients $R_{8}^{{\rm 2\mbox{-}loop}\,   (2k)}$ 
maintain the space-time parity and cyclicity.

For comparison of the results at strong coupling and at  two loops,
we introduce rescaled remainder functions  \cite{Brandhuber:2009da}.
For the $AdS_{3}$ $2n$-gon, they are  defined by
\eqb\label{Rbar}
\bar{R}_{2n} := \frac{R_{2n}-R_{2n,{\rm reg}}}{
R_{2n,{\rm reg}}-(n-2)R_{6,{\rm reg}}}
\comma
\eqe
at strong coupling, and a similar expression at  two loops, 
where $R_{2n,{\rm reg}}$ stands for the remainder function for 
the regular $2n$-gon.
Since     $R_{2n}$, $R_{2n}^{\rm 2\mbox{-}loop} $  reduce 
to superpositions of the contributions from $(n-2)$ regular hexagons 
in the  low-temperature/collinear limit 
\cite{Alday:2009yn,Heslop:2010kq}, 
the rescaled remainder functions are calibrated to take $-1$ in this limit.
It has been observed  numerically \cite{Brandhuber:2009da,DelDuca:2010zp} 
that  $\bar{R}_{8}$ at strong coupling and 
$\bar{R}_{8}^{\rm 2\mbox{-}loop}$
at  two loops are very similar.

Given  (\ref{R8}) and (\ref{Oct2L}), we are now able to
derive analytic expansions of $\bar{R}_{8}$ and 
$\bar{R}^{\rm 2\mbox{-}loop}_{8}$. 
By noting that  the remainder functions for the regular hexagon and octagon are
\eqb
R_{6, {\rm reg}} =\frac{7\pi}{12} , \quad 
R_{8, {\rm reg}} =R_{8}^{(0)} ,
\eqe
and 
\eqb
R_{6, {\rm reg}}^{\rm 2\mbox{-}loop}= -\frac{\pi^{4}}{36} , \quad 
R_{8, {\rm reg}}^{\rm 2\mbox{-}loop} 
=R^{\rm 2\mbox{-}loop \, (0)}_{8},
\eqe
respectively, we obtain 
\eqb
\bar{R}_{8} = \sum_{k=1}^{\infty} \bar{R}^{(2k)}_{8}(\phi) l^{2k},
\qquad 
\bar{R}_{8}^{\rm 2\mbox{-}loop} 
= \sum_{k=1}^{\infty} \bar{R}^{{\rm 2\mbox{-}loop} \, (2k)}_{8}(\phi) l^{2k}
\eqe
where the first few coefficients at strong coupling are
\eqb
\bar{R}^{(2)}_{8}(\phi)\Eqn{=}
 \frac{\frac{1}{8\pi}-\frac{\log 2}{16}}{\frac{\pi}{12}-\frac{\log^2 2}{2}} \approx
 -0.1637687, \nn \\
\bar{R}^{(4)}_{8}(\phi) 
\Eqn{\approx} 0.0174868 + 0.000667828\cos 4\phi, \\
\bar{R}^{(6)}_{8}(\phi)
\Eqn{\approx}  -0.00160021 - 0.000173979 \cos 4\phi, \nn
\eqe
whereas those at  two loops are
\eqb
\bar{R}^{\rm 2\mbox{-}loop(2)}_{8}(\phi)
\Eqn{=}\frac{\log 2-1}{4\log^2 2}\approx-0.15966848, \nn\\
\bar{R}^{\rm 2\mbox{-}loop(4)}_{8}(\phi)
\Eqn{\approx}0.0163067 + 0.00118658\cos 4\phi, \\
\bar{R}^{\rm 2\mbox{-}loop(6)}_{8}(\phi)
\Eqn{\approx}-0.00141679 - 0.00029145 \cos 4\phi. \nn
\eqe
We observe that they are indeed close to each other (but different).

\subsection{Decagon}

Let us move on to a discussion on the $AdS_{3}$ decagon.
In this case, 
the analytic expression of the remainder function 
in \cite{Heslop:2010kq,Gaiotto:2010fk} is 
\eqb\label{R10-2L}
R_{10}^{\rm 2\mbox{-}loop}= -\frac{\pi^{4}}{12} -\frac{1}{2} 
\sum_{k=1}^{10} \log(u_{k}) \log(u_{k+1}) \log(u_{k+2}) \log(u_{k+3})
\comma
\eqe
with $u_{k}=u_{k+10}$. 
The cross-ratios $u_{k}$ are related to $c_{13}^{\pm}, c_{14}^{\pm}$ by\footnote{
We identify $x_{k}^{\pm}$ in \cite{Heslop:2010kq} with $x_{k}^{\mp}$, so that
the $\bbZ_{10}$ symmetry from the parity and cyclicity matches
at strong coupling and at two loops.
}
\eqb\label{uc}
\begin{array}{ll} 
{ \displaystyle
u_{10} = \frac{1+c_{13}^{+}}{1+c_{13}^{+} +c_{14}^{+}}
\comma } 
  &
   { \displaystyle  
   u_{1}  = \frac{1+c_{13}^{-}}{1+c_{13}^{-} +c_{14}^{-}}
}
\comma  \\
 { \displaystyle
u_{2}  = \frac{c^{+}_{14}}{c^{+}_{14}+1}
\comma 
} & 
 { \displaystyle
 u_{3} = \frac{c^{-}_{14}}{c^{-}_{14}+1}
\comma 
} \\
 { \displaystyle
u_{4}  = \frac{1+c_{13}^{+}+ c_{14}^{+}}{(1+c_{13}^{+})(1 +c_{14}^{+})}
\comma } &
 { \displaystyle
 u_{5}  =  \frac{1+c_{13}^{-}+ c_{14}^{-}}{(1+c_{13}^{-})(1 +c_{14}^{-})}
\comma } \\
 { \displaystyle
u_{6} = \frac{c^{+}_{13}}{c^{+}_{13}+1}
\comma } &
 { \displaystyle
 u_{7}  = \frac{c^{-}_{13}}{c^{-}_{13}+1}
\comma } \\
 { \displaystyle
u_{8}  = \frac{1+c_{14}^{+}}{1+c_{13}^{+} +c_{14}^{+}}
\comma } &
{ \displaystyle
 u_{9}  =  \frac{1+c_{14}^{-}}{1+c_{13}^{-} +c_{14}^{-}}  \period
} 
\end{array}
\eqe

 Since $R_{10}^{\rm 2\mbox{-}loop}$ is invariant under the symmetries
associated with the space-time parity and cyclicity, its  high-temperature 
expansion is largely constrained similarly to $\Delta A_{\rm BDS}$.
Consequently,  by substituting 
the cross-ratios (\ref{eq:c13}), (\ref{eq:c14}) into  (\ref{R10-2L}), (\ref{uc}), one obtains 
the following expansion of the remainder function at  two loops: 
\eqb\label{Deca2L}
R_{10}^{\rm 2\mbox{-}loop} = \sum_{k=0}^\infty R_{10}^{{\rm 2\mbox{-}loop} \, (k)}  \, l^{2k/5}
\comma
\eqe
where the first few coefficients are 
\eqb
R_{10}^{\rm 2\mbox{-}loop \, (0)} 
 \Eqn{=} -\frac{\pi^{4}}{12} -5 \log^{4}\bigl( 2 \cos\frac{\pi}{5} \bigr) , \nn \\
R_{10}^{ \rm 2\mbox{-}loop \, (1)} \Eqn{=}  R_{10}^{ \rm 2\mbox{-}loop \, (2)} 
= R_{10}^{\rm 2\mbox{-}loop \, (3)} = 0 , \\
R_{10}^{\rm 2\mbox{-}loop \, (4)} \Eqn{=} 
D_{2}  \cdot 
\bigl\vert 
 Y^{(2)} (\tilde{M}_1e^{i\varphi_{1}},\tilde{M}_2 e^{i\varphi_{2}} )
 \bigr\vert^{2} , \nn 
\eqe
with 
\eqb\label{D2}
 D_{2} \Eqn{=} 
 2^{4} \sqrt{5}
\cos^{6}\Bigl(\frac{ 2\pi}{5} \Bigr)
\log^{2}\Bigl( 2\cos \frac{\pi}{5}\Bigr)
\Bigl[3\sqrt{5} -2^{4}  \cos^{2}\Bigl(\frac{ \pi}{5} \Bigr)
\log\Bigl( 2\cos \frac{\pi}{5}\Bigr) \Bigr] \period 
\eqe
One can check that $ R_{10}^{\rm 2\mbox{-}loop \,(0)}$ 
agrees with the numerical value of $R_{10}^{\rm 2\mbox{-}loop}$ 
for the regular decagon \cite{Brandhuber:2009da}.
The structure of the expansion is very similar to that of $R_{10}$ at strong coupling,
which is  understood as  due to  the space-time symmetries
(see Appendix D for details.).

\begin{figure}[tbp]
\begin{center}
\resizebox{80mm}{!}{\includegraphics{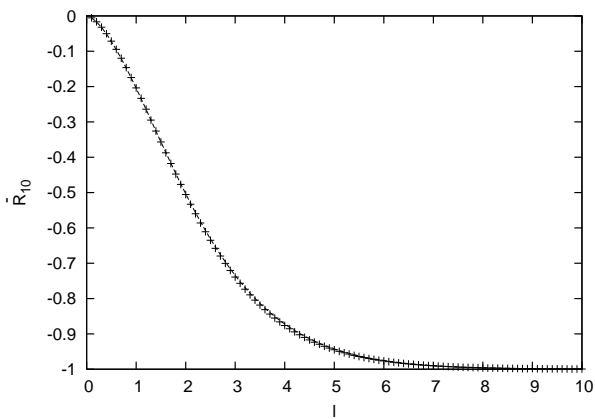}}
\end{center}
\caption{Plots of the $l$-dependence of 
the rescaled remainder functions at strong coupling (points)
and at two loops (dashed lines). The functions are evaluated at
$\tilde{M}_1=\tilde{M_2}=1$ and $\varphi_1=\varphi_2={\pi/20}$.}
\label{fig4}
\end{figure}

Given the above result,  we can  compare the rescaled remainder functions
at strong coupling and at  two loops. 
From   (\ref{Rbar})  and a similar expression with  
\eqb
R_{10, {\rm reg}}  =R_{10}^{(0)} \comma \quad
R_{10, {\rm reg}}^{\rm 2\mbox{-}loop} =R_{10}^{\rm 2\mbox{-}loop\, (0)} \comma
\eqe
we find  that 
\eqb
\bar{R}_{10}  \Eqn{=}  \bar{C}_{8/5}\,
  \big\vert Y^{(2)} (\tilde{M}_1e^{i\varphi_{1}},\tilde{M}_2 e^{i\varphi_{2}} )
  \big\vert^{2}
 \cdot  l^{8/5} + \calO(l^{12/5}) 
\comma \nn \\
\bar{R}_{10}^{\rm 2\mbox{-}loop} \Eqn{=}  \bar{C}_{8/5}^{\rm2\mbox{-}loop}
 \big\vert Y^{(2)} (\tilde{M}_1e^{i\varphi_{1}},\tilde{M}_2 e^{i\varphi_{2}} )
  \big\vert^{2}  \cdot l^{8/5}
+ \calO(l^{12/5}) \comma
\eqe
where
\eqb
  \bar{C}_{\frac{8}{5}} \Eqn{=} \frac{-\frac{1}{5} \tan\frac{\pi}{5}+B_{2}}{
\frac{\pi}{5}-\frac{5}{2}\log^{2}\bigl(2\cos\frac{\pi}{5}\bigr)} \approx 
-0.0441916 \comma \nn \\
 \bar{C}_{\frac{8}{5}}^{\rm2\mbox{-}loop} \Eqn{=}
 \frac{-D_{2}}{5\log^{4}\bigl(2\cos\frac{\pi}{5}\bigr)}
 \approx  -0.0449039 \period
\eqe
$B_{2}$ and $D_{2}$ are given  in (\ref{eq:B_2}) and  (\ref{D2}),
respectively. 
Again, we observe  that they  are very close.
 We note that the two functions are also very close for finite $l$
(Fig.~\ref{fig4}).
This suggests that 
 not only the high-temperature expansion but also the remainder function 
itself is  strongly constrained by the Y-system and the space-time symmetries,
in addition to the collinear limits  \cite{Alday:2010ku,Heslop:2010kq,Gaiotto:2010fk}.

\section{Conclusions and discussion}

In this paper we have studied the remainder functions of the gluon
scattering amplitudes at strong coupling by using the integrable bulk and 
boundary perturbation of conformal field theory.
In particular we have studied the minimal surfaces in $AdS_3$, which 
correspond to the Wilson loops with a $2\tilde{n}$-sided light-like 
polygonal boundary.

The minimal surface is described by the TBA system, and 
the related integrable model is the homogeneous 
sine-Gordon model with purely imaginary
resonance parameters.
This model is obtained by the integrable perturbation of 
generalized parafermions.
We have investigated high-temperature (small mass) expansion of 
the free energy, 
the T-functions and Y-functions
of this model, which give the remainder function around the kinematic
configurations associated with regular polygons.
The high-temperature expansion of the free energy is calculated by the
bulk perturbation of the CFT.
For the T-functions,
we have introduced the $g$-functions whose ratios obey the same 
integral equations and asymptotic conditions. 
By using this relation, 
we have calculated the T-functions.

For the 8-point amplitudes (octagon), the relevant CFT is the critical
Ising model.
Since we know the exact $g$-function in this case, we have performed 
all-order high-temperature expansion of the remainder function.
We have compared this result with the 
2-loop remainder function, and have observed that 
the two results show similar power series expansions
with very close coefficients.

We have  also been able to obtain an explicit formula for 
the first order correction to the remainder function
in the case of the 10-point amplitudes (decagon).
The correction agrees with the numerical 
solution of the TBA equations for small masses.
We have compared this result with the proposed 2-loop remainder function.
Again, we have observed that the rescaled remainder functions
have similar power series structures
with close coefficients.
This observed similarity suggests that 
their power series structure is 
strongly constrained by the Y-system and the space-time symmetries,
in addition to the collinear limits
\cite{Alday:2010ku,Heslop:2010kq,Gaiotto:2010fk}.

The Y-functions are
obtained from the cross-ratios of the T-functions.
A notable observation in our discussion is that the T- and Y-functions
in the CFT limit is given by the modular S-matrix through the $g$-functions.
This suggests an interesting relation between the modular S-matrices and
solutions of constant Y-systems, which are used to compute
the central charge of CFT using the dilogarithm
identities\cite{Kirillov:1987,Nahm:1992sx,Kuniba:2010ir}.
Although we have also observed that the same integral equations are derived
from the $g$- and T-functions, the role of the boundary perturbation of
CFT in the context of gluon
scattering amplitudes is not quite clear
at this moment. These points would deserve further investigations.

For future direction, it would be possible to extend our discussion to the
cases
of more than 10-point amplitudes, amplitudes with more general kinematic
configurations corresponding to $AdS_{4}$ and $AdS_{5}$, and form
factors\cite{Maldacena:2010kp}.
For these purposes, it would be important
to understand multi-parameter integrable deformations of the generalized
parafermionic CFT.
In addition, the underlying integrable models/CFTs for the $AdS_{5}$ case
are yet to be clarified. Taking into account the case of the $AdS_{5}$
hexagon\cite{Alday:2009dv,Hatsuda:2010vr},  one may expect them to be obtained
by some deformation from the $AdS_{4}$ case\cite{Hatsuda:2010cc}.
Details should, however, be discussed further.

Regarding higher order expansions, the approach adopted in this paper
requires higher correlation functions in the presence of both the bulk and
the boundary deformations. 
In order to cover the full kinematic region of gluon momenta, one needs
higher order expansion, which is connected to the
low-temperature (large mass)
region\cite{Alday:2009dv,Hatsuda:2010cc,Alday:2010ku,Bartels:2010ej}.
On the other hand, a different approach to study
the analytic expansion of the T-functions
has been given by Bazhanov et al.~for
kink (massless) TBA systems (see Appendix C)
\cite{Bazhanov:1994ft,Bazhanov:1996dr}.
For massive systems, see for example \cite{Bazhanov:1996aq, Fioravanti:2003kx}.
It would be interesting to apply this
to  the study of the minimal surface,
as well as to understand the relation between these two approaches.

\vspace{3ex}

\begin{center}
{\bf Acknowledgments}
\end{center}

We would like to thank  J.~Suzuki and Z.~Tsuboi 
for useful discussions and conversations, and R.~Tateo
for useful discussions and for pointing out
the relation between the T- and $g$-functions to us.
The work of K.~S. and Y.~S. is supported in part by Grant-in-Aid
for Scientific Research from the Japan Ministry of Education, Culture, 
Sports, Science and Technology. The work of K.~S. is also supported by
Keio Gijuku Academic Development Funds.

\vspace{3ex}

\appendix

\begin{center}
{\bf Appendices}
\end{center}

\section{Expansion of $F(l,\varphi)$}\label{sec:deri-oct}

In this appendix, we derive the high-temperature expansion of $F(l,\varphi)$ defined by \eqref{eq:F(l,phi)}.
 First we expand $F(l,\varphi)$ in $\sin\varphi$, 
\begin{align}
F(l,\varphi)&=\frac{\cos\varphi}{\pi}\int_0^\infty dt \frac{\cosh t}{\cosh^2 t-\sin^2 \varphi}\log(1+e^{-l \cosh t}), \nn \\
&=\frac{\cos\varphi}{\pi}\sum_{n=0}^\infty \sin^{2n}\varphi\, \Phi_{2n+1}(l),
\label{eq:F-exp0}
\end{align}
where 
\begin{align}
\Phi_{m}(l)\equiv \int_0^\infty \frac{dt}{\cosh^m t}\log(1+e^{-l \cosh t})
=\sum_{k=1}^\infty \frac{(-1)^{k-1}}{k} \int_0^\infty \frac{dt}{\cosh^m t} e^{-kl\cosh t}.
\label{eq:Phi_m-0}
\end{align}
It is easy to see that the $m$-th derivative of $\Phi_m(l)$ is expressed in terms of the modified Bessel function of the second kind,
\begin{align}
\Phi_m^{(m)}(l)=\sum_{k=1}^\infty (-1)^{k+m-1} k^{m-1} K_0(kl).
\label{eq:Phi_m^m}
\end{align}
The summation \eqref{eq:Phi_m^m} for $m=2n+1$ can be evaluated by using the following formula \cite{GR:1980},
\begin{align}
\sum_{k=1}^{\infty} (-1)^{k}\cos(ka) K_{0}(kl)&\nn\\ 
  = \frac{1}{2} \left( \gamma_{E} + \log \frac{l}{4\pi} \right) 
  &+ \frac{\pi}{2} \sum_{j=1}^{\infty} 
         \left[ \frac{1}{ \sqrt{l^2 +[(2j-1) \pi-a]^2} }  
          -\frac{1}{2\pi j}\right] \nonumber \\
 &+ \frac{\pi}{2} \sum_{j=1}^{\infty} 
    \left[ \frac{1}{ \sqrt{l^2 +[(2j-1) \pi+a]^2} }  -\frac{1}{2\pi j} \right].
\end{align}
For example,
\begin{align}
\Phi_3^{(3)}(l)&=-\pi \sum_{j=1}^\infty \frac{-l^2+2(2j-1)^2\pi^2}{[l^2+(2j-1)^2\pi^2]^{5/2}},\label{eq:Phi_3^3}\\
\Phi_5^{(5)}(l)&=3\pi \sum_{j=1}^\infty \frac{3l^4-24(2j-1)^2\pi^2l^2+8(2j-1)^4\pi^4}{[l^2+(2j-1)^2\pi^2]^{9/2}}.\label{eq:Phi_5^5}
\end{align}
By integrating both sides
in \eqref{eq:Phi_3^3}, \eqref{eq:Phi_5^5} etc.,
we find the general structure
\begin{align}
\Phi_{2n+1}'(l)=\sum_{j=1}^{2n}\frac{\Phi_{2n+1}^{(j)}(0)}{(j-1)!}l^{j-1}
+(-1)^n \pi \sum_{j=1}^\infty \frac{l^{2n}}{(2j-1)^{2n}\pi^{2n}\sqrt{l^2+(2j-1)^2\pi^2}}.
\end{align}
One can explicitly check this for small $n$'s.
Therefore
\begin{align}
&\Phi_{2n+1}(l) 
=\sum_{j=0}^{2n}\frac{\Phi_{2n+1}^{(j)}(0)}{j!}l^{j}  \nn \\ 
&\hspace{1.0cm}+(-1)^n \pi \sum_{m=n}^\infty \begin{pmatrix} -\frac{1}{2} \\ m-n \end{pmatrix}
\(1-\frac{1}{2^{2m+1}}\)\zeta(2m+1)\frac{1}{2m+1}\(\frac{l}{\pi}\)^{2m+1}  .
\label{eq:Phi_2n+1}
\end{align}
The remaining task is to determine $\Phi_{2n+1}^{(j)}(0)$ ($j=0,1,\dots,2n$).
If we define $f(x)=\log(1+e^{-x})$, from \eqref{eq:Phi_m-0} we find
\begin{align}
\Phi_{2n+1}^{(j)}(0)=f^{(j)}(0)\int_0^\infty \frac{dt}{\cosh^{2n-j+1}t}
=\frac{2^{2n-j-1}\Gamma(n-\frac{j-1}{2})^2}{\Gamma(2n-j+1)}f^{(j)}(0).
\label{eq:J_2n+1(0)}
\end{align}
From the following series expansion,
\begin{align}\label{logcosh}
\log \cosh \frac{x}{2}
=\sum_{j=2}^\infty \frac{(2^{j}-1)B_j}{j!\cdot j}x^j,
\end{align}
we get
\begin{align}
f(0)=\log 2,\qquad f^{(j)}(0)=\frac{(2^j-1)B_j}{j} \;\;(j \geq 1) ,
\label{eq:f^j(0)}
\end{align}
where the Bernoulli numbers are defined by
\begin{equation}
\frac{x}{e^x-1}=\sum_{n=0}^\infty \frac{B_n}{n!}x^n.
\end{equation}
Substituting \eqref{eq:J_2n+1(0)} and \eqref{eq:f^j(0)} into \eqref{eq:Phi_2n+1}, 
we obtain the expansion of $\Phi_{2n+1}(l)$.
In summary, the expansions of $\Phi_{2n+1}(l)$ are 
\begin{align}
\Phi_1(l)&=\frac{\pi}{2}\log 2-\frac{l}{2}\(\log\frac{1}{l}+1+\log\pi-\gamma_{\rm E}\)\notag \\
&\hspace{0.5cm}+\pi\sum_{m=1}^\infty \begin{pmatrix}-\frac{1}{2}\\ m\end{pmatrix}\(
1-\frac{1}{2^{2m+1}}\)\zeta(2m+1)\frac{1}{2m+1}\(\frac{l}{\pi}\)^{2m+1},
\label{eq:J_1-exp}\\
\Phi_{2n+1}(l)&=\frac{2^{2n-1}\Gamma(n+\frac{1}{2})^2}{\Gamma(2n+1)}\log2
+\sum_{j=1}^{2n}\frac{2^{2n-1}(1-2^{-j})}{j\cdot j!}\frac{\Gamma(n-\frac{j-1}{2})^2B_j}{\Gamma(2n-j+1)}l^j\notag \\
&\hspace{0.5cm}+(-1)^n \pi \sum_{m=n}^\infty \begin{pmatrix} -\frac{1}{2} \\ m-n \end{pmatrix}
\(1-\frac{1}{2^{2m+1}}\)\zeta(2m+1)\frac{1}{2m+1}\(\frac{l}{\pi}\)^{2m+1} .
\label{eq:J_2n+1-exp}
\end{align}
Substituting \eqref{eq:J_1-exp} and \eqref{eq:J_2n+1-exp} into \eqref{eq:F-exp0}, we obtain
\begin{align}
F(l,\varphi)
=\frac{1}{2}\log 2 -\frac{l}{2\pi}\biggl[ \cos\varphi\(\log\frac{1}{l}+1+\log\pi-\gamma_{\rm E}\)+\varphi \sin \varphi\biggr]+\sum_{p=2}^\infty c_p(\varphi) l^{p}.
\end{align}
The expressions of $c_p(\varphi)$ are given by
\begin{align}
&c_{2k}(\varphi)\nn
=\frac{(2^{2k}-1)B_{2k}}{4k(2k)!}\sin^{2k}\varphi ,\\
&c_{2k+1}(\varphi)\nn\\
&=\frac{\cos\varphi}{\pi}\sum_{n=0}^k \sin^{2n}\varphi\cdot (-1)^n \pi 
\begin{pmatrix} -\frac{1}{2} \\ k-n \end{pmatrix}\(1-\frac{1}{2^{2k+1}}\)\zeta(2k+1)\frac{1}{2k+1}
\frac{1}{\pi^{2k+1}} \nonumber \\
&=\begin{pmatrix} -\frac{1}{2} \\ k \end{pmatrix}
\frac{1}{(2k+1)\pi^{2k+1}}\(1-\frac{1}{2^{2k+1}}\)\zeta(2k+1)\cos\varphi\;{}_2F_1(-k,1;\frac{1}{2}-k;\sin^2\varphi) .
\end{align}
Note that the sum over even $p$ can be performed,
\begin{align}
\sum_{k=1}^\infty c_{2k}(\varphi)l^{2k}
=\frac{1}{4}\(l\sin\varphi+2\log\frac{1+e^{-l\sin\varphi}}{2}\).
\end{align}
Thus we finally arrive at the expansion \eqref{eq:F-exp1}.

\section{Perturbation of the $(\grp{SU}(2)_{-1/2} \times \grp{SU}(2)_1)/\grp{SU}(2)_{1/2}$ coset
model}\label{sec:coset-pert}

In this appendix, we compute the high-temperature expansion of the free energy in 
the non-unitary $(\grp{SU}(2)_{-1/2} \times \grp{SU}(2)_1)/\grp{SU}(2)_{1/2}$ coset model perturbed by 
the primary field $\phi_{1,1,3}$ with dimension $\Delta=\bar{\Delta}=1/5$.
This model plays an important role in analyzing the remainder function for the decagon with $M_1=M_2$.
This model is equivalent to the non-unitary minimal model ${\cal M}_{3,5}$ perturbed by the relevant operator $\Phi=\Phi_{1,3}$,
whose action takes the following form,
\begin{align}
S=S_{\rm CFT}+\hat{\lambda} \int\!d^2x\, \Phi(x).
\end{align}
Using the result in \cite{Fateev:1993av}, we can write down the coupling-mass relation
\begin{align}
\hat{\lambda}&=\hat{\kappa} M^{8/5},\nn \\
\hat{\kappa}^2&=\frac{1}{\pi^2}\gamma\(-\frac{1}{5}\)\gamma\(\frac{3}{5}\)
\left[\frac{\sqrt{\pi}}{8}\gamma\(\frac{1}{4}\)\right]^{16/5}.
\end{align}
Note that $\hat{\kappa}$ is purely imaginary in this case as well as in
the scaling Lee-Yang model ${\cal M}_{2,5}$.
The central charge of the UV CFT is $c=-3/5$, and the ground state corresponds to the operator
$\Phi_{1,2}$ with dimension $\Delta_0=\bar{\Delta}_0=-1/20$.
Thus the effective central charge is given by
\begin{align}
\hat{c}=c-12(\Delta_0+\bar{\Delta}_0)=\frac{3}{5},
\end{align}
which is one-half of the central charge for the $\grp{SU}(3)_2/\grp{U}(1)^2$ coset CFT as expected.

Let us consider the free energy of this model.
Near the high-temperature limit $l\to 0$, the free energy is expanded as
\begin{align}
\hat{F}(l)=\frac{\pi}{6}\hat{c}+\frac{1}{4}l^2+\sum_{n=1}^\infty \hat{f}^{(n)}l^{8n/5}.
\end{align}
where
\begin{align}
\hat{f}^{(n)}&=\frac{\pi}{6}\hat{\kappa}^n \hat{C}^{(n)},\\
\hat{C}^{(n)}&=12\frac{(-1)^n}{n!}(2\pi)^{2\Delta-1}\int \prod_{j=1}^{n-1} \frac{d^2z_j}{(2\pi |z_j|)^{2(1-\Delta)}}\notag \\
&\hspace{1cm}\times
\langle \Phi_0(\infty,\infty)\Phi(1,1)\Phi(z_1,\bar{z}_1)\cdots\Phi(z_{n-1},\bar{z}_{n-1}) \Phi_0(0,0)\rangle_{\rm connected}. 
\end{align}
Recall that the vacuum operator is $\Phi_0=\Phi_{1,2}$ and the perturbing operator is $\Phi=\Phi_{1,3}$.
The first non-vanishing coefficient is $\hat{C}^{(1)}$:
\begin{align}
\hat{C}^{(1)}=-12(2\pi)^{-3/5}C_{\Phi_0\Phi\Phi_0},
\end{align}
where $C_{\Phi_0\Phi\Phi_0}$ is the structure constant, which was computed in \cite{Dotsenko:1985hi},
\begin{align}
(C_{\Phi_0\Phi\Phi_0})^2=(D^{(1,2)}_{(1,3)(1,2)})^2=\gamma\(-\frac{1}{5}\)\gamma\(\frac{3}{5}\)\gamma\(\frac{4}{5}\)^2 .
\end{align}
Thus the leading correction is
\begin{align}
\hat{f}^{(1)}&=\frac{1}{16}\(\frac{\pi}{4}\)^{1/5}\gamma\(\frac{1}{4}\)^{8/5}\gamma\(-\frac{1}{5}\)
\gamma\(\frac{3}{5}\)\gamma\(\frac{4}{5}\) .
\end{align}
Returning to the discussion on the decagon, we obtain an expansion of the free energy with $M_1=M_2=M$,
\begin{align}
A_{\rm free}^{\rm decagon}|_{M_1=M_2}=2\hat{F}(l)=\frac{\pi}{5}+\frac{1}{2}l^2+2\hat{f}^{(1)}l^{8/5}+\cdots.
\label{eq:free-M1=M2}
\end{align}

\section{Generalization to complex masses}\label{sec:complex-mass}

Here we discuss how to extend the expansions of the Y-functions to the case with complex masses.
We focus on the decagonal case $\tilde{n}=5$.
However the generalization to $\tilde{n}\geq 6$ is straightforward.

Let us consider the expansion of the Y-functions.
From the quasi-periodicity (\ref{Yperiod}) and analyticity, 
the Y-functions should have the following expansion \cite{Zamolodchikov:1991et}
\begin{align}
Y_j(\theta)=\frac{1}{2}\sum_{k=-\infty}^\infty Y_j^{(k)}e^{\frac{2k\theta}{5}},
\end{align}  
with $Y_2^{(k)}=(-1)^k Y_1^{(k)}$.
In addition, the reality condition  (\ref{reality})  constrains the coefficients as
$Y_j^{(-k)}=\overline{Y_j^{(k)}}$, which gives
\eqb\label{Y12}
Y_{1}(\theta) \Eqn{=} \half \sum_{k=0}^{\infty} \Bigr(Y^{(k)}
e^{\frac{2k}{5} \theta}
   + \overline{Y^{(k)}} e^{-\frac{2k}{5} \theta} \Bigr)
  \comma\nn \\
 Y_{2}(\theta) \Eqn{=} \half \sum_{k=0}^{\infty} (-1)^{k} 
  \Bigl(Y^{(k)} e^{\frac{2k}{5} \theta}
   + \overline{Y^{(k)}} e^{-\frac{2k}{5} \theta} \Bigr)
  \period
\eqe
The coefficients $Y^{(k)}$ are functions of $m_j$ and $\bar{m}_j$, 
and $Y^{(2)}$ here coincides with \eqref{eq:exp_Y} if all the masses are real.
$Y^{(k)}$ are expanded in powers of 
$l^{2/5}$, with 
the leading behavior $Y^{(k)} \sim l^{2k/5}$ for small $l$ 
\cite{Zamolodchikov:1991et,Dorey:1997rb}.
Moreover, according to \cite{Dorey:1999cj}, this leading behavior
is thought of as coming from the boundary perturbation in (\ref{Gexpansion}).
The form of the expansion (\ref{Gexpansion}) then implies that
the subleading corrections are given in powers of $l^{4/5}$. Thus,
one may have
\eqb\label{Y(k)}
 Y^{(k)} = \sum_{p =0}^{\infty} b_{k,2p} \,  l^{\frac{2}{5}(k+2p)}
 \comma
\eqe
where $l=ML$ and $M$ is an overall mass scale.
This is  in accord with the double expansion in terms of 
$ l e^{\pm \theta} $ discussed in \cite{Dorey:1997rb}. 
At low orders, the absence of terms of order $l^{2(k+p')/5}$ with odd $p'$
is also confirmed from the Y-system by  
following \cite{Zamolodchikov:1991vx}.
We have checked that the above expansion is consistent 
with numerical results.

It is convenient here 
to write the first few terms of the expansion of $Y_1(\theta)$ in $l^{2/5}$,
\begin{align}
2Y_1(\theta)=2b_{00}+(b_{20}e^{\frac{4\theta}{5}}+\bar{b}_{20}e^{-\frac{4\theta}{5}})l^{4/5}+{\cal O}(l^{6/5}) ,
\label{eq:Y_1-exp}
\end{align}
where we have used the fact that $b_{00}=\bar{b}_{00}=2\cos(\pi/5)$ 
and $b_{10}=b_{02}=b_{12}=0$ as seen in Appendix D.
The coefficients $b_{k,2p}$ depend on both $\tilde{M}_je^{-i\varphi_j}$ and $\tilde{M}_je^{i\varphi_j}$ in general.
It is important to notice that from \eqref{TBAeqs} the TBA equations for $Y_j(\theta)=\tilde{Y}_j(\theta-i\varphi_j)$ are given by
\begin{align}
\log Y_{j}(\theta)
&=-\frac{1}{2}(\bar{m}_j e^{\theta}+m_j e^{-\theta})+K * \log (1+Y_{j-1})(1+Y_{j+1}).
\end{align}
In order to reveal the complex mass dependence of $b_{k,2n}$, we consider the decoupling limit (chiral limit) $l \to 0$.
In this limit, the new functions $Y_j^{\rm kink}(\theta)=Y_j(\theta-\log(l/2))$ satisfy the kink TBA equations 
(see \cite{Zamolodchikov:1989cf} for example)
\begin{align}
\log Y_{j}^{\rm kink}(\theta)=-\tilde{M}_je^{-i\varphi_j} e^{\theta}+K * \log (1+Y_{j-1}^{\rm kink})(1+Y_{j+1}^{\rm kink}).
\label{eq:kinkTBA}
\end{align}
From the periodicity, $Y_{j}^{\rm kink}(\theta)$ have the expansion,
\begin{align}
Y_{j}^{\rm kink}(\theta)=\frac{1}{2}\sum_{k=0}^\infty Y_{j}^{{\rm kink}(k)}e^{\frac{2k\theta}{5}} .
\label{eq:exp-Y_j^kink}
\end{align}
\eqref{eq:kinkTBA} suggests that the coefficients $Y_{j}^{{\rm kink}(k)}$ are functions of 
$\tilde{M}_je^{-i\varphi_j}$, not of $\tilde{M}_je^{i\varphi_j}$.
On the other hand, by taking the decoupling limit in 
\eqref{Y12} with \eqref{Y(k)}, we obtain
\begin{align}
Y_1^{\rm kink}(\theta)=\lim_{l\to0}Y_1\(\theta-\log\frac{l}{2}\)=b_{00}
+\frac{1}{2}\sum_{k=0}^\infty 2^{\frac{2k}{5}}b_{k0}e^{\frac{2k\theta}{5}}.
\label{eq:exp-Y_1^kink}
\end{align}
Comparing \eqref{eq:exp-Y_1^kink} with \eqref{eq:exp-Y_j^kink}, we obtain
\begin{align}
Y_{1}^{{\rm kink}(0)}=2b_{00},\quad Y_{1}^{{\rm kink}(k)}=2^{2k/5}b_{k0}.
\end{align} 
These relations show that the coefficients $b_{k0}$ ($k\geq 1$) depend on $\tilde{M}_je^{-i\varphi_j}$, 
not on $\tilde{M}_je^{i\varphi_j}$.
Similarly, $\bar{b}_{k0}$ ($k\geq 1$) are functions of $\tilde{M}_je^{i\varphi_j}$.
In summary, we can get $b_{k0}$ ($\bar{b}_{k0}$) for the complex masses 
by replacing $\tilde{M}_j \to \tilde{M}_je^{-i\varphi_j}$ ($\tilde{M}_je^{i\varphi_j}$) in
$b_{k0}$ for the real masses assuming the analyticity in $\tilde{M}_{j}$.
However, $b_{k,2n}$ ($n\geq 1$) are functions of $\tilde{M}_je^{-i\varphi_j}$ and $\tilde{M}_je^{i\varphi_j}$,
and 
the above argument does not applied to them.
We already know the small $l$ expansion of the Y-functions for the real masses up to order $l^{4/5}$ (see \eqref{eq:Y-exp-real}).
The coefficient of $l^{4/5}$ is
\begin{align}
b_{20}^{\rm real}=Y^{(2)}(\tilde{M}_1,\tilde{M}_2).
\end{align}
Using this result, we can obtain $b_{20}$ and $\bar{b}_{20}$ for the complex masses by the above prescription,
\begin{align}
b_{20}^{\rm complex}
&=Y^{(2)}(\tilde{M}_1e^{-i\varphi_1},\tilde{M}_2e^{-i\varphi_2}), \\
\bar{b}_{20}^{\rm complex}
&=Y^{(2)}(\tilde{M}_1e^{i\varphi_1},\tilde{M}_2e^{i\varphi_2}).
\end{align}
Substituting these equations into \eqref{eq:Y_1-exp}, we obtain
the small $l$ expansion of the Y-functions for the complex masses as in \eqref{eq:exp-Y}.

\section{Structure of  expansions at  higher orders}\label{sec:HOrder}

In the main text, we have obtained the first order high-temperature
expansion of the Y-functions  for the decagon
by using the conformal perturbation.  
In this appendix, we show that  the structure of
the  high-temperature expansion is largely constrained by 
the Y-system, the structure of the conformal perturbation, 
and the symmetries associated with the space-time parity and cyclicity,
although one still needs 
higher order perturbations to find precise values of the coefficients. 
In the following, we concentrate on the case of the $AdS_{3}$ decagon, 
but the discussion below can be extended to more general cases. 

We start with the expansion (\ref{Y12}), and  (\ref{Y(k)}) which is in 
accord with the conformal perturbation as discussed in Appendix C.
Substituting these into the
Y-system (\ref{Ysystem}) for the $AdS_{3}$ decagon, one obtains 
a double expansion in $e^{2 \theta/5}$ and $l^{2/5}$,
in which each coefficient should vanish.
For the first few orders, we then find, e.g., 
\eqb
&&  
b_{00} = 2\cos\bigl(\frac{\pi}{5}\bigr) \comma \quad 
b_{10} =b_{02} =b_{12}=0 \comma \nn \\
&&  
 b_{04} = \frac{1}{5} \sin^{2}\bigl( \frac{2\pi}{5}\bigr) 
 \cdot  b_{20} \bbar_{20}
\comma \quad 
b_{40} = \frac{2}{5} \sin^{2}\bigl( \frac{\pi}{5}\bigr) \cdot (b_{20})^{2}  
\period
\eqe
The expansion of the Y-functions in turn gives the expansions
of the cross-ratios and $\Delta A_{\rm BDS}$. 
Using the relations among $b_{k,2p}$ obtained
from the Y-system, we find that
\eqb
\Delta A_{\rm BDS} = \sum_{k=0}^{\infty} A_{k} l^{2k/5} \comma
\eqe
with 
\eqb
&& A_{0} = -\frac{2}{5} \log^{2}\bigr(2\cos \frac{\pi}{5}\bigr) \comma \quad
 A_{4} = B_{2} \cdot b_{20}  \bbar_{20} \comma \\
&& A_{6} = B_{2} 
 ( b_{20} \bbar_{22} +  b_{22}\bbar_{20}
+  b_{30} \bbar_{30} ) 
-40 \cos^{4}\bigl( \frac{2}{5}\pi\bigr) \cdot b_{30} \bbar_{30} 
\comma \nn
\eqe
and $A_{1}=A_{2}=A_{3}=A_{5}=A_{7}=0$. $B_{2}$ is given in (\ref{eq:B_2}).
The result shows that one can obtain the expansion of $\Delta A_{\rm BDS}$
up to $\calO(l^{12/5})$ once $b_{20}$ is known. In the main text and Appendix C,
$b_{20}$ is found   to be 
 $ b_{20} = Y^{(2)} \bigl(\tilde{M}_1e^{-i\varphi_{1}},\tilde{M}_2 
 e^{-i\varphi_{2}} \bigr) $. 

One can argue that 
the absence of $A_{k}$ with odd  $k$ is understood as
a consequence of the $\bbZ_{10} $ symmetry 
due to the space-time parity and cyclicity, 
$x_{i}^{-} \to x_{i+1}^{+}, x_{i}^{+} \to x_{i}^{-}$.
The $\bbZ_{10}$ transformation is concisely expressed by 
the  $Y$-functions as \cite{Gaiotto:2010fk}
\eqb\label{Z10}
  Y_{j}(\theta) \to Y_{j}\bigl(\theta+\frac{\pi}{2} i\bigr)  \comma
\eqe
or in terms of the expansion coefficients and  the TBA masses, 
$
  Y^{(k)} \to e^{\frac{\pi i }{5} k} Y^{(k)} $ and 
   $ m_{j} \to m_{j}/i $, respectively. 
The cross-ratios $c_{13}^{\pm}, c_{14}^{\pm}$ transform as
$c_{13}^{-} \to c_{24}^{+}$, $c_{14}^{-} \to c_{25}^{+}$, $c^{+}_{13} \to c_{13}^{-}$, 
$c_{14}^{+} \to c_{14}^{-}$, where $c_{24}^{+}, c_{25}^{+}$ are given in
(\ref{c-others}).
$\Delta A_{\rm BDS}$ in   (\ref{DBDS10}) indeed has this symmetry.  
Note also  that $A_{k}$ consists of terms of the form 
$\prod  b_{p_{j}, 2q_{j}} \prod  \bbar_{\bar{p}_{j}, 2\bar{q}_{j}} $ with
$k=\sum (p_{j}+2q_{j}) + \sum (\bar{p}_{j}+2\bar{q}_{j})$, which 
transform under (\ref{Z10}) as $ \prod  b_{p_{j}, 2q_{j}} 
\prod  \bbar_{\bar{p}_{j}, 2\bar{q}_{j}} 
\to e^{\frac{\pi i}{5} \Delta p} \prod  b_{p_{j}, 2q_{j}} \prod  
\bbar_{\bar{p}_{j}, 2\bar{q}_{j}} $
with $\Delta p:= \sum p_{j} -\sum \bar{p}_{j}$. Thus, 
unless non-trivial cancellations with other terms occur,
which is unlikely because $b_{p,2q}$ are functions of $m_{j},\bar{m}_{j}$,
the terms with $\Delta p \neq 0$ (mod $10$) are projected out 
by successive actions of the $\bbZ_{10}$ transformation. 
For odd $k$, $\Delta p$ necessary becomes non-zero and hence 
$A_{k}$ is projected out. This leads to the  symmetry  $l^{2/5} \to -l^{2/5}$.
 Furthermore, the  $\bbZ_{10}$ symmetry is then
promoted to a continuous symmetry 
$Y^{(k)} \to e^{i\phi k} Y^{(k)}$, which  in turn corresponds 
to a continuous imaginary shift of $\theta$
in $Y_{j}(\theta)$
and hence the world-sheet rotational symmetry.

$\Delta A_{\rm BDS}$ is also invariant under the space-time parity symmetry:
$x_{i}^{+} \to x_{i}^{-}$ with  
the order of the labeling of the cusps  being reversed. In the case of the
decagon, this gives rise to $c_{13}^{\pm} \leftrightarrow c_{14}^{\pm}$,
which is equivalent to $l^{2/5} \to -l^{2/5}$ and
$Y^{(k)} \leftrightarrow \overline{Y^{(k)}}$. 
Combined with  the above symmetry $l^{2/5} \to -l^{2/5}$,
the parity results in the symmetry 
$Y^{(k)} \leftrightarrow \overline{Y^{(k)}}$. 
Consequently, 
only the terms with $\Delta p=0$ and maintaining the symmetry
$Y^{(k)} \leftrightarrow \overline{Y^{(k)}}$, such as
$b_{p,2q} \bbar_{p,2q'} + \bbar_{p,2q} b_{p,2q'} $, 
are allowed in the expansion.

In addition, since the remainder function has to have the parity and
$\bbZ_{10}$ symmetries, $A_{\rm period} + A_{\rm free} (=R_{10} 
-  \Delta A_{\rm BDS} )$ 
also maintains these symmetries. Oppositely, since 
$A_{\rm period} + A_{\rm free}$ is invariant under the  world-sheet rotational
symmetry by definition,  so is the remainder function.

Regarding the functional form of $b_{p,2q}$,
the conformal perturbation and the expansion in 
$l e^{\pm \theta}$ \cite{Dorey:1997rb}  for complex TBA masses 
suggest that $b_{p,2q}$ are given by summation of  terms of the from
$ l^{-\frac{2}{5}(p+2q)} \bar{m}_{j_{1}}^{2/5} \cdots 
\bar{m}_{j_{p+q}}^{2/5} \cdot m_{j'_{1}}^{2/5}  \cdots  m_{j'_{q}}^{2/5}  $.
Indeed, given this form, one finds that $Y^{(k)}$  transform  as
$Y^{(k)} \to e^{\frac{\pi i}{5}k} Y^{(k)}$ and
$Y^{(k)} \to e^{-i\frac{2k}{5}\varphi} Y^{(k)}$
under the  $\bbZ_{10}$ transformation
$m_{j} \to m_{j}/i$ and the world-sheet rotation equivalent to
$m_{j} \to e^{i\varphi} m_{j}$, respectively. This is in accord with 
(\ref{Z10}) and the
above argument.
Note that the invariants under the latter symmetry depend
on  the phases of  $m_{j}$ only through
their deference $\varphi_{12} = \varphi_{1} -\varphi_{2}$.

Our results of the expansion are all  consistent with the above arguments. 
However, further investigations are needed for definite conclusions.

We can also  study the structure of the  high-temperature expansion of 
the remainder function at  two loops. From    the expansion 
of $Y_{j}(\theta)$ in  (\ref{Y12}),  (\ref{Y(k)}), we find that the remainder function
(\ref{R10-2L}) is expanded as
\eqb
R_{10}^{\mbox{\scriptsize 2-loop}} 
= \sum_{k=0}^{\infty} R_{10}^{{\rm 2\mbox{-}loop} \, (k)}   l^{2k/5}
\comma  
\eqe
with
\eqb
&& R_{10}^{{\rm 2\mbox{-}loop} \, (0)} 
= -\frac{\pi^{4}}{12} -5 \log^{4}\bigl( 2 \cos\frac{\pi}{5} \bigr)
\comma \quad 
R_{10}^{{\rm 2\mbox{-}loop} \, (4)} = D_{2} \cdot b_{20}\bbar_{20}
\comma \\
&& 
R_{10}^{\rm 2\mbox{-}loop \,(6)} = 
     D_{2} (b_{20} \bar{b}_{22} + b_{22}\bar{b}_{20} + b_{30}\bar{b}_{30})
     + 5 \cdot 2^{8} \cos^{9}\Bigl(\frac{2 \pi}{5} \Bigr)
      \log^{2}\Bigl( 2 \cos \frac{\pi}{5}\Bigr) \cdot b_{30}\bar{b}_{30}
      \comma \nn
\eqe
and  $R_{10}^{{\rm 2\mbox{-}loop} \, (1)} = R_{10}^{{\rm 2\mbox{-}loop} \, (2)} 
= R_{10}^{{\rm 2\mbox{-}loop} \, (3)}  =R_{10}^{{\rm 2\mbox{-}loop} \, (5)}  =
R_{10}^{{\rm 2\mbox{-}loop} \, (7)}  =0$. $D_{2}$ is given in (\ref{D2}).
Thus, the structure of the expansion is very similar to that of $\Delta A_{\rm BDS}$.
This is because $R_{10}^{\mbox{\scriptsize 2-loop}} $
has the parity and $\bbZ_{10}$ symmetries and hence the structure of
the expansion is  strongly constrained as in the case of $\Delta A_{\rm BDS}$.

\vspace{3ex}

%
%
\def\thebibliography#1{\list
{[\arabic{enumi}]}{\settowidth\labelwidth{[#1]}\leftmargin\labelwidth
\advance\leftmargin\labelsep
\usecounter{enumi}}
\def\newblock{\hskip .11em plus .33em minus .07em}
\sloppy\clubpenalty4000\widowpenalty4000
\sfcode`\.=1000\relax}
\let\endthebibliography=\endlist
\vspace{3ex}
\begin{center}
{\bf References}
\end{center}
\par \vspace*{-2ex}

\end{document}